\documentclass[a4paper,11t]{article}
\usepackage[left=2.54cm, right=2.54cm, top=2.54cm, bottom=2.54cm]{geometry}
\usepackage{rotating}
\usepackage{multirow}
\usepackage{tikz}
\usepackage{booktabs}
\usepackage{amsmath}
\usepackage{amsfonts}
\usepackage{amssymb}

\usepackage{epstopdf}
\usepackage[caption=false]{subfig}
\usepackage[numbers,sort&compress]{natbib}
\usepackage[colorlinks=true, allcolors=blue]{hyperref}

\begin{document}

\begin{center}
Hypothesis testing for two population means: parametric or non-parametric test?   
\vskip 1cm
M. Tsagris$^a$, A. Alenazi$^b$, K.~M. Verrou$^c$ and N. Pandis$^d$ 
\vskip 0.5cm
$^a$ Department of Economics, University of Crete, Rethymnon, Greece, \\ \href{mailto:mtsagris@uoc.gr}{mtsagris@uoc.gr} \\ ksa99ksa99@hotmail.com
$^b$ Department of Mathematics, Northern Border University, Arar, Saudi Arabia, \\ \href{mailto:ksa99ksa99@hotmail.com}{ksa99ksa99@hotmail.com} \\ 
$^c$ Department of Computer Science, University of Crete, Herakleion, Greece, \\
\href{mailto:kleioverrou@yahoo.com}{kleioverrou@yahoo.com} \\ 
$^d$ Department of Orthodontics and Dentofacial Orthopedics, University of Bern, Bern, Switzerland, \\ 
\href{mailto:npandis@yahoo.com}{npandis@yahoo.com}
\vskip 1cm

\textbf{Abstract} \\
\end{center}
The parametric Welch $t$-test and the non-parametric Wilcoxon-Mann-Whitney test are the most commonly used two independent sample means tests. More recent testing approaches include the non-parametric, empirical likelihood and exponential empirical likelihood. However, the applicability of these non-parametric likelihood testing procedures is limited partially because of their tendency to inflate the type I error in small sized samples. In order to circumvent the type I error problem, we propose simple calibrations using the $t$ distribution and bootstrapping. The two non-parametric likelihood testing procedures, with and without those calibrations, are then compared against the Wilcoxon-Mann-Whitney test and the Welch $t$-test. The comparisons are implemented via extensive Monte Carlo simulations on the grounds of type I error and power in small/medium sized samples generated from various non-normal populations. The simulation studies clearly demonstrate that a) the $t$ calibration improves the type I error of the empirical likelihood, b) bootstrap calibration improves the type I error of both non-parametric likelihoods, c) the Welch $t$-test with or without bootstrap calibration attains the type I error and produces similar levels of power with the former testing procedures, and d) the Wilcoxon-Mann-Whitney test produces inflated type I error while the computation of an exact p-value is not feasible in the presence of ties with discrete data. Further, an application to real gene expression data illustrates the computational high cost and thus the impracticality of the non parametric likelihoods. Overall, the Welch t-test, which is highly computationally efficient and readily interpretable, is shown to be the best method when testing equality of two population means. \\
\\
\textbf{keywords}: Sections; Hypothesis testing; Welch $t$-test; empirical likelihood; bootstrap \\
\\
\textbf{amscode}: 62G10; 62F40

\section{Introduction}
Hypothesis testing for two population (univariate) means has been approached with a plethora of tests over the years \citep{brunner2000, fagerland2009, fagerland2012, derrick2016, chen2016, dwivedi2017, sudhir2018}. The most commonly used methods are the Welch $t$-test that relaxes the (restrictive and possibly unrealistic) assumption of equal variances imposed by the student's $t$-test and its non-parametric alternative Wilcoxon-Mann-Whitney (WMW) test. More recently developed approaches include the (non-parametric) empirical likelihood (EL) \citep{owen1988, owen1990, owen2001} and the exponential empirical likelihood (EEL), also known as non-parametric tilting or exponential tilting \citep{efron1981}. 

It is known, however, that the WMW, EL and EEL tests have limitations. For instance, EL and EEL inflate the type I error when applied to small sized samples, in both univariate and multivariate population means \citep{diciccio1991, qin1994, tanizaki2004, emerson2009a, emerson2009b, tsagris2017}. \citep{tsao2001, tsao2004, emerson2009b} addressed this issue, by proposing a small sample calibration for the EL, applicable to the one population mean test\footnote{For a list of small sample improvements (in terms of the estimated probability of type I error) in the case of one population mean see \cite{emerson2009a}.} only. \cite{jing1995} proposed a Bartlett correction\footnote{The 3rd and 4th cumulants of the test statistic are $O(n^{-3})$ and $O(n^{-4})$ respectively.} of the EL for the two population means problem. The correction led to more accurate estimates of the type I error, but the simulation study was rather small and narrow. Deriving the Bartlett correction can be challenging and therefore should be estimated via bootstrap \citep{hall1990}, and with an increased computational cost. On the same spirit, \cite{amaral2004, namba2004, bartolucci2007, amaral2010, tsagris2017} proposed bootstrap calibration\footnote{A great advantage of bootstrap is the absence of any parametric assumptions about the data or the test statistic.} of the EL test statistic directly. In contrast to EL, the EEL, does not admit Bartlett correction \citep{jing1996} and an adjustment was suggested by \cite{zhu2008}. This should not be considered as a disadvantage, according to \cite{schennach2007}, as the Bartlett correction is sometimes ineffective, while EEL shares the same higher order asymptotic properties of the EL \cite{schennach2007} as established in \cite{newey2004}. Finally, the  non-parametric WMW test is based on ranks and hence unable to capture the differences in the means. The statistic of the WMW without ties is asymptotically normally distributed and although exact p-value computation is possible it becomes difficult as the number of combinations increase. In the presence of ties in small sample sizes, asymptotic normality does not hold and exact p-value computation is not feasible.

The parametric Welch $t$-test is one of the oldest and most recognized test as it is always included in introductory statistics textbooks. However, its value has not been adequately recognized mainly due to the unrealistic normality assumption. If the normality assumption is not satisfied, non-parametric testing procedures are suggested. In reality, the Welch test statistic asymptotically follows the normal distribution and not the data per se. The Welch $t$-test utilizes the sample means which in conjunction with the central limit theorem strengthens the asymptotic normality assumption of its test statistic. 

A theoretical comparison of the parametric and non-parametric methods for two population means is not feasible with small sample sizes, and possibly of no practical interest for practitioners and applied researcher, especially when different distributions have to be examined. In addition, it is not possible to examine the theoretical properties of those tests for all pairs of existing distributions and theory and practice do not always meet. EL for instance is asymptotically correct, but in practice it is not size correct with small sample sizes. 

What is missing from the literature is not another theoretical comparison of the Welch, WMW, EL and EEL tests, but a solid large scale Monte Carlo study that compares them under various distributional scenarios. Similar to the Monte Carlo simulations of \cite{tsagris2017} who compared the multivariate versions of these tests for constrained non Gaussian data. Their simulations provided evidence that the multivariate $t$-test (James test) with bootstrap calibration consistently estimated the type I error adequately, in contrast to its non-parametric counterparts EL and EEL. Despite revealing some practical benefits of the James test, their simulations were rather "narrow" in the sense that they used only a few distributional scenarios. 

Compared to \cite{tsagris2017} the present study is wider in scope, aiming to provide evidence supporting the use of the Welch $t$-test for hypothesis testing of two univariate population means via a large scale Monte Carlo simulation study. The simulation studies presented in this paper consider various asymmetric and bimodal distributions, distributions with limited support (e.g. Beta and von Mises distribution), and discrete distributions. Following \cite{tsagris2017}, this paper proposes to calibrate the EL and EEL against a $t$ distribution, and not against the $\chi^2$ distribution. The results clearly indicate that calibration of the EL test statistic with a $t$ distribution corrects the inflated type I error. Bootstrap on the other hand, improves the estimated type I error of the EL, EEL and Welch $t$-test. Among the equally performing testing procedures (EL, EEL and Welch $t$-test), the Welch $t$-test is considered the most favorable test, due to its accurate performance, its simplicity and its extremely low computation cost, that increases only slightly when bootstrap calibration is applied. Computational efficiency has not yet received requisite attention, despite being a highly essential feature in the era of massive and big data. In the field of bioinformatics for example, gene expression data commonly number $55,000$ variables. In such cases, applying a time inefficient non-parametric test that does not produce more accurate results than the Welch $t$-test is impractical and undesirable. 

The structure of this paper is as follows. At first two classical tests, the Welch $t$-test and the WMW test are presented. Then the EL and EEL test statistics and their relationship with the Welch $t$-test are presented in Sections \ref{empirical} and \ref{exponential}. Two calibrations of those test statistics are presented in Section \ref{calibrations}. In Section \ref{comparisons} Monte Carlo simulations compare the probability of type I error and the corresponding power of those testing procedures. The computational advantage of the Welch $t$-test is illustrated by comparing the computational cost of the testing procedures using 4 real gene expression datasets in Section \ref{genes}. Finally, the conclusions and findings of the paper are summarised in Section \ref{conclusions}.

\section{Classical tests for two population means}
Suppose there are collections of observations from two populations $(x_1, \ldots,x_{n_1})$ and $(y_1, \ldots,y_{n_2})$, where $n_x$ and $n_y$ denote the two sample sizes, are available, and that the aim is to make inference about their population means.

\subsection{Welch $t$-test} \label{welch}
\cite{welch1947} proposed a test for linear form of hypotheses of population means when the population variances are unknown and not assumed equal. The test statistic specifically for the two samples $t$-test is
\begin{eqnarray} \label{tw}
T_w=\frac{\bar{x}-\bar{y} }{\sqrt{\frac{s_x^2}{n_x}+\frac{s_y^2}{n_y}}},
\end{eqnarray} 
where $\bar{x}$ and $\bar{y}$ denote the two sample means and $s_x^2$ and $s_y^2$ are the two sample variances. Under $H_0$, $T_w \sim t_{\nu}$, with $t_{\nu}$ denoting the $t$ distribution with $\nu$ degrees of freedom and $\nu$ is given by \citep{satterthwaite1946, welch1947}
\begin{eqnarray}  \label{nu}
\nu \simeq \frac{\left(\frac{s_x^2}{n_x} + \frac{s_y^2}{n_y}\right)^2}{\frac{s_x^4}{n_x^2(n_x-1)} + \frac{s_y^4}{n_y^2(n_y-1)}}.
\end{eqnarray}

\subsection{Wilcoxon-Mann-Whitney rank based test}
\cite{wilcoxon1945} proposed the Wilcoxon rank-sum test as a non-parametric, rank based, alternative to the $t$-test. An adjusted test was also proposed, independently, by \cite{mann1947} and for this it is also known as Wilcoxon-Mann–Whitney (WMW) test. The procedure for WMW as proposed by \cite{wilcoxon1945} consists of the following steps. \begin{enumerate}
\item Merge the observations from both samples in one vector ${\bf u} = ({\bf x}, {\bf y})$ and rank the observations.
\item Calculate $W = \sum_{i=1}^{n_x}R_{xi}$, where $R_{xi}$ is rank of the $i$-th observation of the first sample. 
\end{enumerate}

The statistical software R \citep{R2018} computes $W - \frac{n_x(n_x+1)}{2}$ and the standardised test statistic is given by 
\begin{eqnarray} \label{mwz}
Z = \frac{W - \frac{n_x(n_x+1)}{2} - 0.5n_xn_y}{\sqrt{\frac{n_xn_y(n_x+n_y+1)}{12}}}.
\end{eqnarray}
If there are ties in the ranks, the denominator of (\ref{mwz}) should be corrected as follows
\begin{eqnarray} \label{mwzc}
Z_c = \frac{W - \frac{n_x(n_x+1)}{2} - 0.5n_xn_y}{\sqrt{\frac{n_xn_y}{12}\left(n_x+n_y+1 -\sum_{i=1}^k\frac{t_i^3-t_i}{(n_x+n_y)(n_x+n_y-1)}\right)}},
\end{eqnarray}
where $t_i$ is the number of observations sharing rank $i$, and $k$ is the number of (distinct) ranks. Under $H_0$, asymptotically, $Z$ and $Z_c$ follow the standard normal distribution.   

\section{Non-parametric likelihoods} 

\subsection{Empirical likelihood} \label{empirical}
Empirical likelihood is a non-parametric likelihood developed by \cite{owen1988,owen1990} in order to perform non-parametric hypothesis testing. In the hypothesis testing of two-sample means, positive probabilities $p_{xi}$, $i=1,\ldots,n_x$ and $p_{yi}$, $i=1,\ldots,n_y$ are assigned to the $i$-th observation of each sample, with $\sum_{i=1}^{n_x}p_{xi}=\sum_{i=1}^{n_y}p_{yi}=1$ \cite{jing1995} and \cite{liu2008}.
The log-likelihood ratio test statistic 
\begin{eqnarray} \label{eltest}
\Lambda =2\left[ \sum_{i=1}^{n_x}\log{n_xp_{xi}} + \sum_{i=1}^{n_y}\log{n_yp_{yi}} \right]
\end{eqnarray}
must be maximised under the following constraints:
\begin{eqnarray} \label{constraints}
\sum_{i=1}^{n_x}p_{xi} = \sum_{i=1}^{n_y}p_{yi}=1 \ \ \text{and}  \ \ 
\sum_{i=1}^{n_x}p_{xi}\left(x_i-\mu\right) = \sum_{i=1}^{n_y}p_{yi}\left(y_i-\mu\right)=0,
\end{eqnarray}
where $\mu$ is the population mean, under $H_0$. Introducing Lagrangian parameters $\lambda_x$ and $\lambda_y$ and after some algebra, the $p_{xi}$ and $p_{yi}$ can be rewritten as $p_{xi}=\frac{1}{n_x\left[1+\lambda_x\left(x_i-\mu\right)\right]}$ and $p_{yi}=\frac{1}{n_y\left[1+\lambda_y\left(y_i-\mu\right)\right]}$ respectively. A numerical search is necessary to obtain the $\lambda_x$, $\lambda_y$, the $\mu$ and essentially the $p_{xi}s$ and $p_{yi}s$ that maximise (\ref{eltest}) leading to
\begin{eqnarray} \label{lambda_el}
\Lambda = 
2\sum_{i=1}^{n_x}\log{\left[1+\lambda_x\left(x_i-\mu\right)\right]} + 2\sum_{i=1}^{n_y}\log{\left[1+\lambda_y\left(y_i-\mu\right)\right]},
\end{eqnarray}
which asymptotically becomes 
\begin{eqnarray} \label{ellambda}
\Lambda & = & \frac{n_x\left(\bar{x}-\mu\right)^2}{s^2_x\left(\mu\right)}+\frac{n_y\left(\bar{y}-\mu\right)^2}{s^2_y\left(\mu\right)}+O_{p}\left(n_0^{-1}\right) \nonumber \\ 
\Lambda & = & T^2_x\left(1+\frac{T^2_x-1}{n_x} \right)^{-1} + T^2_y\left(1+\frac{T^2_y-1}{n_y} \right)^{-1} + O_{p}\left(n_0^{-1}\right),
\end{eqnarray}
where $T_x^2$ and $T_y^2$ are the squares of the one sample $t$-test statistic about $\mu$ of each sample and $n_0=\min{\left\lbrace n_x,n_y \right\rbrace}$. When $H_0$ is true, $\Lambda$ (\ref{ellambda}) follows a $\chi^2_1$ \citep{owen2001}, since $s^2\left(\mu\right)\overset{p}{\rightarrow} \sigma^2$, from Slutsky's theorem \citep{casella2002}, where $s^2\left(\mu\right)$ is the sample variance calculated using the population mean and $\sigma^2$ is the population variance. A proof of the asymptotic distribution of the test statistic, for the multi-sample univariate mean hypothesis test, can be found in \cite{owen2001}, but see Appendix \ref{A1} of the present paper for a different approach applicable to 2 or more samples.  

\subsection{Exponential empirical likelihood} \label{exponential}
Exponential empirical likelihood (EEL) was first introduced by \cite{efron1981} to perform a "tilted" version of the bootstrap for the one sample mean hypothesis testing. The same constraints as in (\ref{constraints}) apply, but the resulting probabilities take an exponential form for each sample $p_{xi}=\frac{e^{\lambda_x x_i}}{\sum_{k=1}^{n_x}e^{\lambda_x x_k}}$ and $p_{yi}=\frac{e^{\lambda_y y_i}}{\sum_{k=1}^{n_y}e^{\lambda_y y_k}}$
and hence the zero-sum constraints in (\ref{constraints}) become
\begin{eqnarray} \label{expconstraint2}
\frac{\sum_{i=1}^{n_x}x_ie^{\lambda_x x_i}}{\sum_{k=1}^{n_x}e^{\lambda_x x_k}} =  \frac{\sum_{i=1}^{n_y}y_ie^{\lambda_y y_i}}{\sum_{k=1}^{n_y}e^{\lambda_y y_k}} = \mu
\end{eqnarray}
Similarly to EL, a numerical search is performed to compute the values of $\lambda_x$ and $\lambda_y$, but note that it is not necessary to search for the common mean. These values are then plugged into the log-likelihood ratio test statistic, which takes the following form
\begin{eqnarray} \label{eeltest}
\Lambda &=& -2\sum_{i=1}^{n_x}\log{\left[n_x\left(\sum_{k=1}^{n_1}e^{\lambda_x x_k}\right)^{-1}e^{\lambda_x x_i}\right]} -2\sum_{i=1}^{n_y}\log{\left[n_y\left(\sum_{k=1}^{n_y}e^{\lambda_y y_k}\right)^{-1}e^{\lambda_y y_i}\right]} \nonumber \\
\Lambda &=& \frac{n_x\left(\bar{x}-\mu\right)^2}{s^2_x\left(\bar{x}\right)}+\frac{n_y\left(\bar{y}-\mu\right)^2}{s^2_y\left(\bar{y}\right)}+O_{p}\left(n_0^{-1/2}\right) \nonumber \\
\Lambda &=& T_x^2\left(1-\frac{1}{n_x} \right)^{-1} + T_y^2\left(1-\frac{1}{n_y} \right)^{-1} + O_{p}\left(n_0^{-1/2}\right),
\end{eqnarray}
where $T_x^2$ and $T_y^2$ are the one sample $t$-test statistic from the true common mean of each sample and $s^2_x\left(\bar{x}\right)$ and $s^2_y\left(\bar{y}\right)$ denote the sample biased variance of each sample. Under $H_0$ $\Lambda$ (\ref{eeltest}) follows asymptotically a $\chi^2_1$ distribution, since $s^2\overset{p}{\rightarrow} \sigma^2$, from Slutsky's theorem \citep{casella2002}, where $s^2$ and $\sigma^2$ are the sample and population variance respectively.

\section{Alternative calibrations of the test statistics} \label{calibrations}

\subsection{Calibration of the EL and EEL test statistics using the $t$ distribution}
The EL test statistic (\ref{ellambda}) and the EEL test statistic (\ref{eeltest}) asymptotically have the same form as the James test statistic \cite{james1954} (see Appendix for a proof), which is also asymptotically equivalent to the Welch test statistic. This implies that the asymptotic powers of the EL and of the EEL test statistics are equivalent to that of the Welch $t$-test. The Welch $t$-test is not affected by the distribution of the populations from which the samples come from because it relies on the central limit theorem. The normal distribution is only asymptotically correct, and the test statistic in the Welch $t$-test is calibrated against the $t$ distribution, that is more reliable for small sample sizes, and leads to higher power. Similarly, the EL and EEL test statistics should be calibrated against a corrected $\chi^2$ distribution as suggested by \cite{james1954} and not against the $\chi^2$ distribution. Obtaining the corrected $\chi^2$ distribution is quite complicated and without meaningful losses the $t$ distribution with degrees of freedom given in (\ref{nu}), can be applied. Similar approaches were suggested by \cite{owen2001} (pg. 26) for the one sample univariate mean, \cite{owen2001} (pg. 30) for the one sample multivariate mean, and \cite{tsagris2017} in the two population multivariate means, who use the $F$ instead of the $\chi^2$ distribution. 

\subsection{Bootstrap calibration}
The non-parametric bootstrap procedure that was applied to all testing procedures, except for the WMW, can be described as follows.

\begin{enumerate}
\item Define the test statistic as $T_{obs}$ that is calculated for the available data $(x_1, \ldots, x_{n_x})$ and $(y_1, \ldots, y_{n_y})$ with sample means $\bar{x}$ and $\bar{y}$ and sample variances $s_x^2$ and $s_y^2$.
\item Transform the data under the null hypothesis
$ x^*_i = x_i - \bar{x}  + \hat{\mu}_c \ \ \text{and} \ \ y^*_i = y_i - \bar{y}  + \hat{\mu}_c $, where $\hat{\mu}_c = \left(\frac{n_x}{s_x^2} +  \frac{n_y}{s_y^2}\right)^{-1}\left(\frac{n_x}{s_x^2}\bar{x} + \frac{n_y}{s_y^2}\bar{y}\right)$ is the estimated common mean under the null hypothesis.
\item Generate bootstrap samples ${\bf x}^{*b}$ and ${\bf y}^{*b}$ by sampling with replacement from ${\bf x}^*$ and ${\bf y}^*$ and calculate for each pair of bootstrap samples the test statistic $T_b$.
\item Repeat step 3 $B$ times, acquire $B$ bootstrap test statistics $T_b^1, ..., T_b^B$ and compute the bootstrap p-value by $ \left[ \sum_{i=1}^B I\left(T^i_b>T_{obs}\right)+1\right]/(B+1) $, 
where $I(.)$ is the indicator function.
\end{enumerate}

\subsection{Exact p-value of the WMW test statistic}
\cite{mann1947} proposed an equivalent to (\ref{mwz}) test statistic $M_{n_x,n_y}=\sum_{i=1}^{n_x} I\left(j:x_j<y_i\right)$. Instead of computing the bootstrap p-value of the WMW test statistic, its exact p-value is computed via the probability generating function of $M_{n_x,n_y}$ \citep{van1999}
\begin{eqnarray} \label{mwexact}
\sum_{k=0}^{n_xn_y}\text{Pr}\left(M_{n_x,n_y} = k\right)z^k=\frac{1}{\binom{n_x+n_y}{n_x}}\frac{\prod_{i=n_x+1}^{n_x+n_y}\left(1-z^i\right)}{\prod_{i=1}^{n_y}\left(1-z^i\right)}.
\end{eqnarray}

The R built-in function \textit{wilcox.test} computes the exact p-value, if the samples contain less than 50 values and there are no ties. Otherwise, the normal approximation (\ref{mwz}) is used. 

\section{Simulation studies for the performance of the testing procedures} \label{comparisons}
Throughout the simulation studies, different sample sizes and unequal variances were chosen, with the smallest size sample coming from the population with the largest variance ($n_1 > n_2$, $\sigma_1^2<\sigma_2^2$). For all test statistics except for the WMW, bootstrap calibration was performed with $499$ bootstrap re-samples. Increasing the number of re-samples to $999$ increased the computational cost, dramatically for the EL and EEL, without any inferential benefit. This agrees with the results of \cite{tsagris2017} who used $299$ re-samples. For the non-parametric likelihood procedures, the $t$ distribution calibration was also examined. Table \ref{info} contains the calibration techniques applied to each testing procedure. All testing procedures except WMW have been self-implemented and are available in the R package \textit{Rfast} \citep{Rfast2018}.  

\begin{table}[htp]
\caption{\textbf{Parametric and non-parametric testing procedures and their calibrations used in the simulation studies}. The first column mention the testing procedure and the next columns refer to the calibration applied on each procedure.}
\begin{center}
{\begin{tabular}{c|c|c|c|c|c} \hline \hline
                   & \multicolumn{5}{c}{Calibration}  \\  \hline
Procedure  &  $\chi^2$ &  $t$  & $N(0, 1)$  &  Bootstrap  & Exact p-value  \\  \hline
EL      &  \checkmark  &  \checkmark    &              &  \checkmark  &              \\   \hline
EEL     &  \checkmark  &  \checkmark    &              &  \checkmark  &              \\   \hline
WMW     &              &                &  \checkmark  &              &  \checkmark  \\   \hline
Welch   &              &  \checkmark    &              &  \checkmark  &  \checkmark   \\ \hline \hline
\end{tabular}}
\end{center}
\label{info}
\end{table}

\subsection{The distributional scenarios}
\begin{figure}[htp]
\centering
{\begin{tabular}{ccc}
\includegraphics[scale = 0.45, trim = 40 40 20 0]{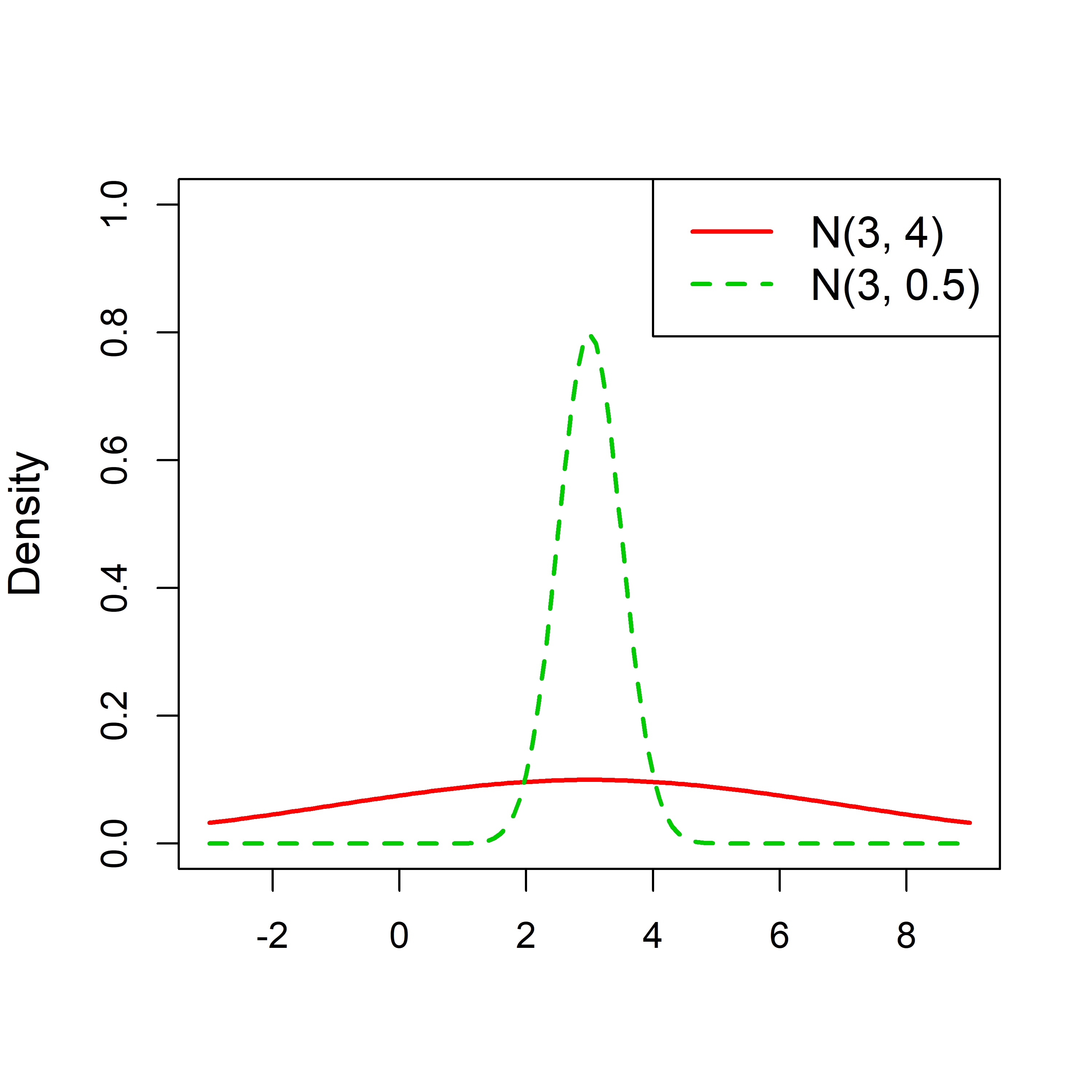} &
\includegraphics[scale = 0.45, trim = 30 40 20 0]{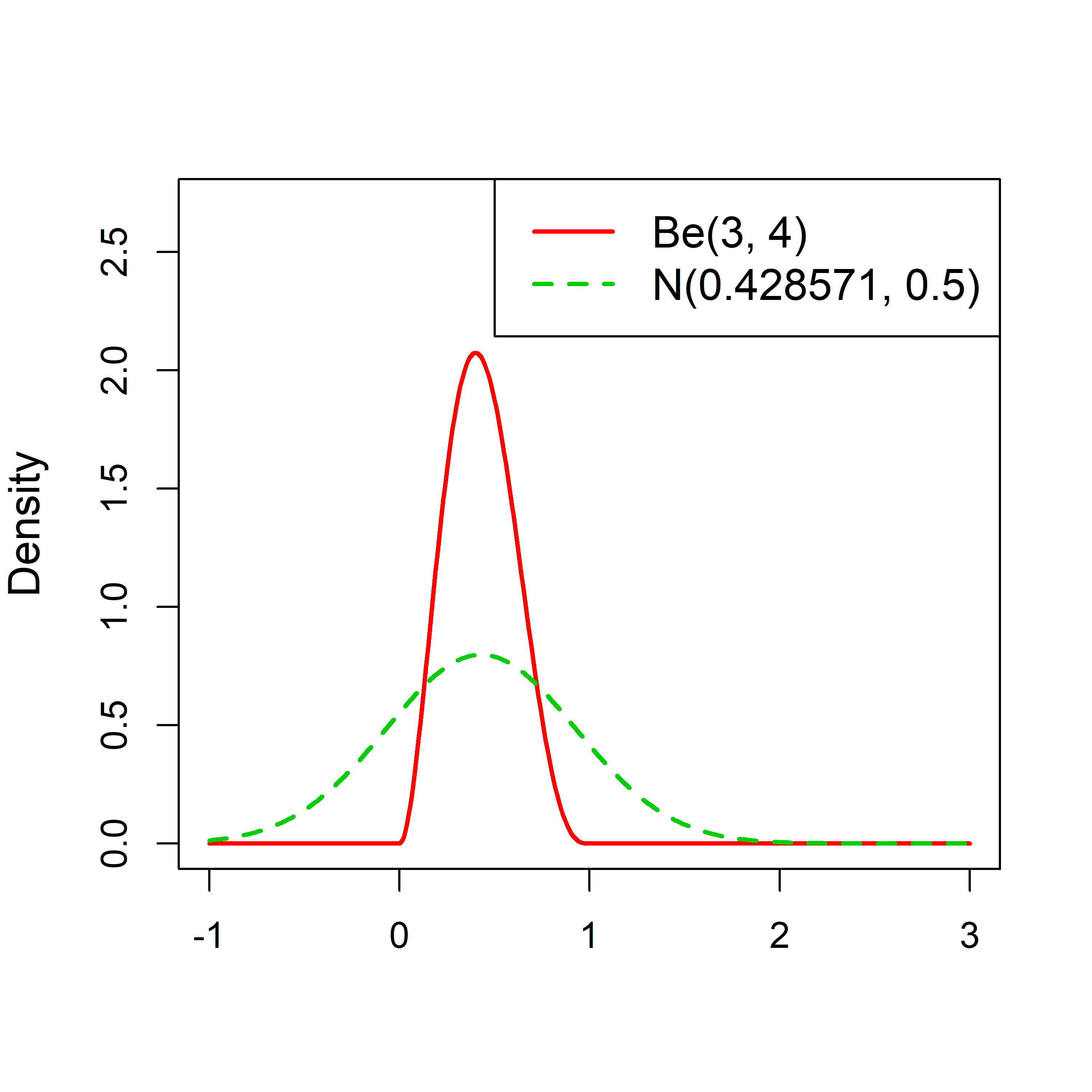} &
\includegraphics[scale = 0.45, trim = 30 40 20 0]{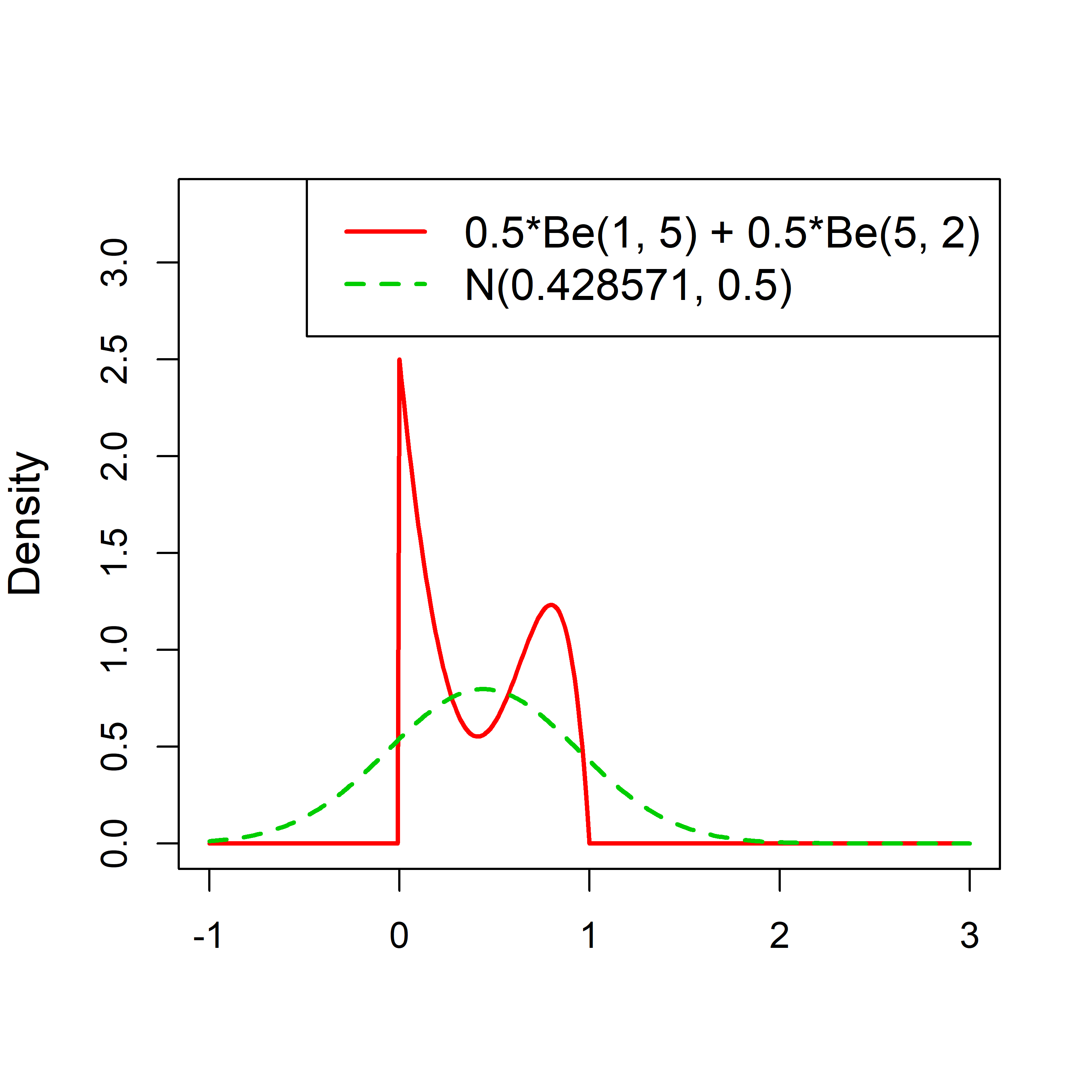}   \\
{\scriptsize (a)}   &   {\scriptsize (b)}  &   {\scriptsize (c)}   \\
\includegraphics[scale = 0.45, trim = 40 40 20 0]{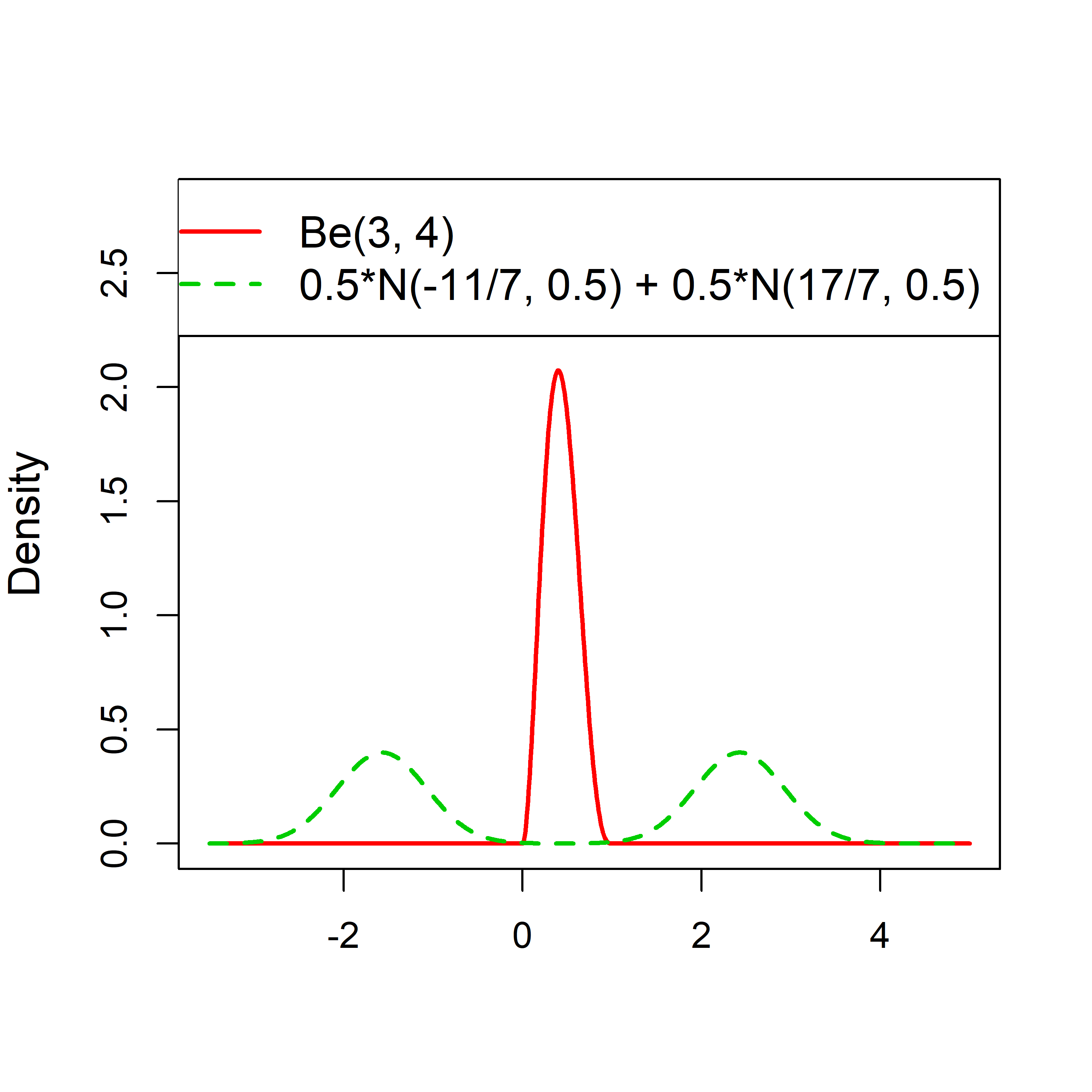} &  
\includegraphics[scale = 0.45, trim = 30 40 20 0]{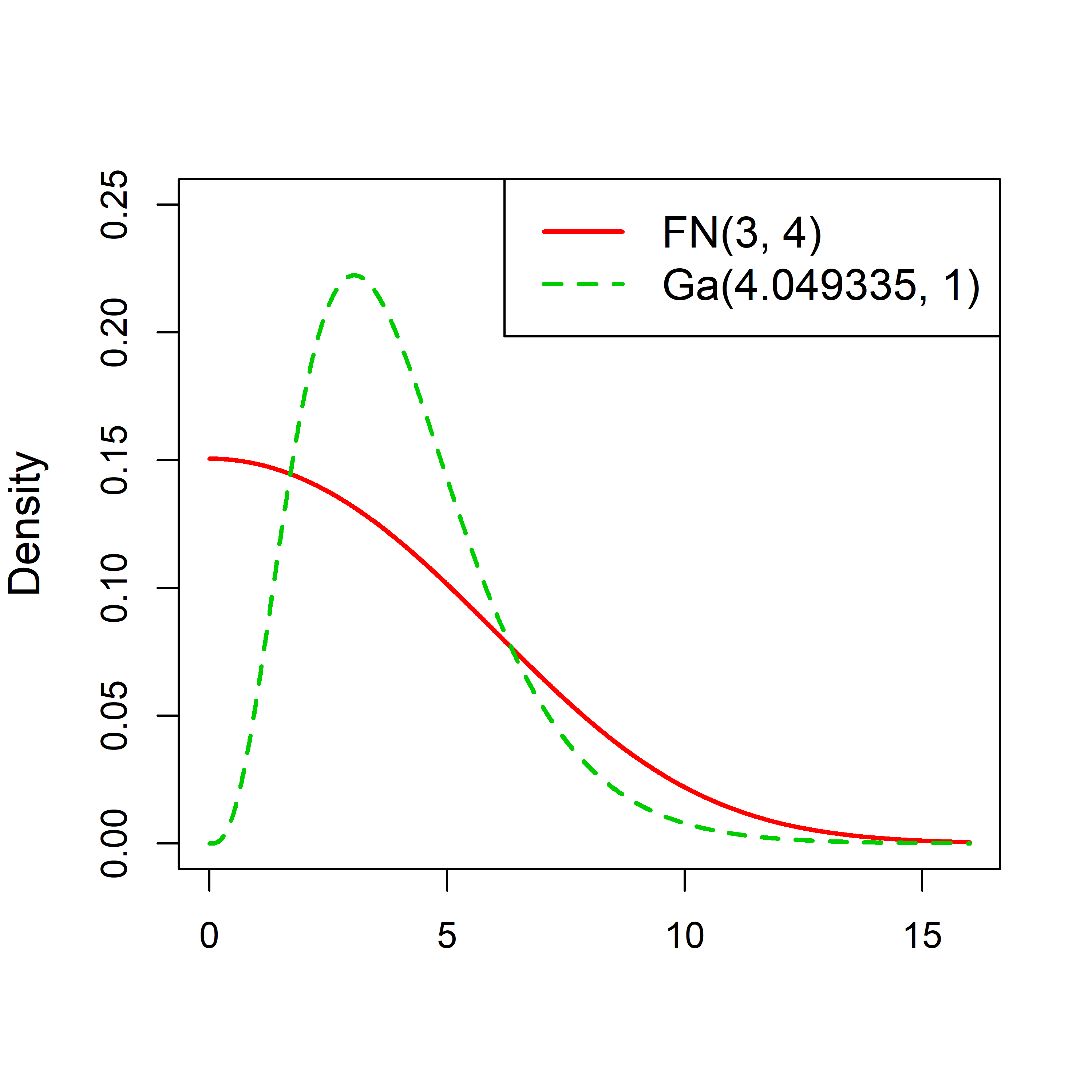}  &  
\includegraphics[scale = 0.45, trim = 30 40 20 0]{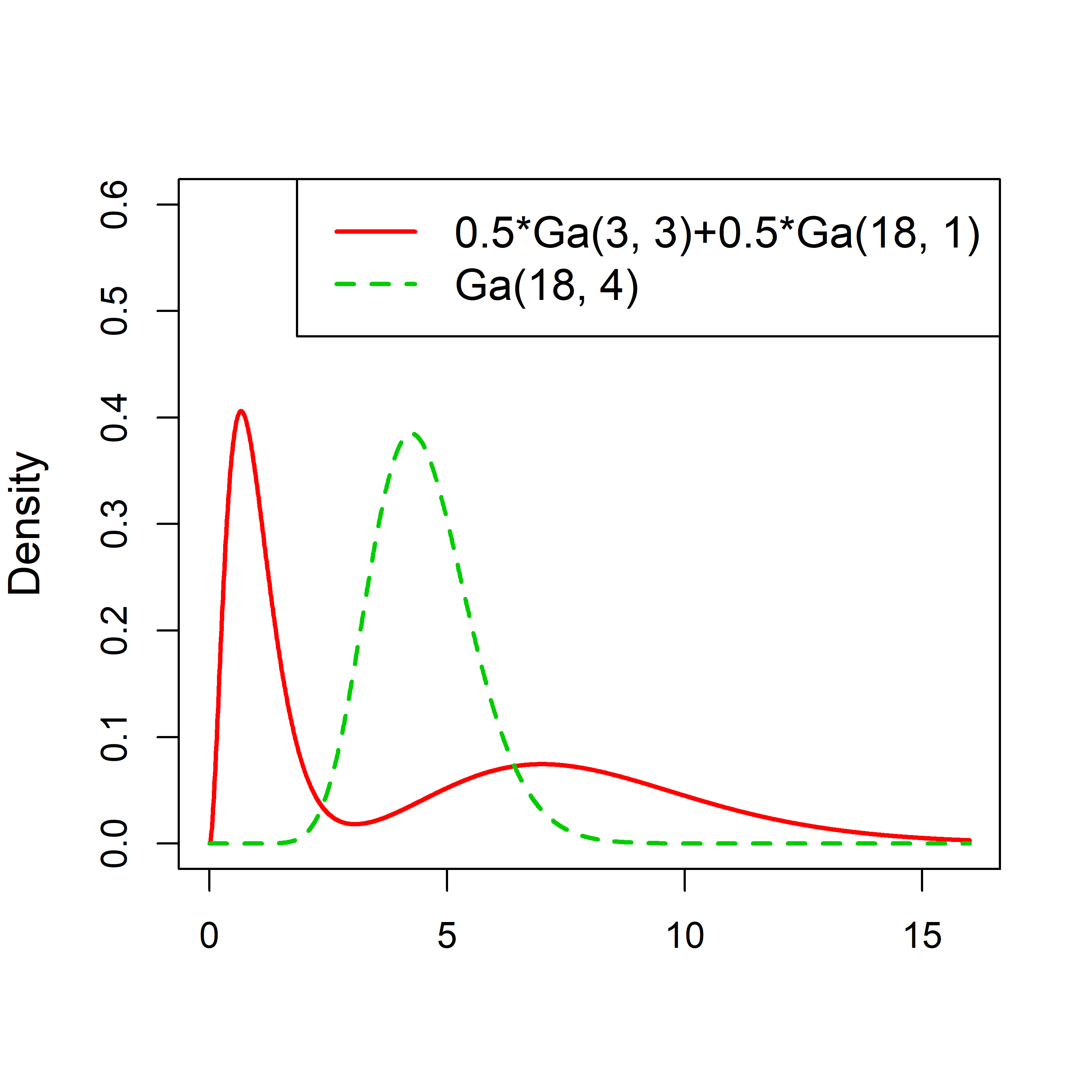}   \\
{\scriptsize (d)}   &   {\scriptsize (e)}  &   {\scriptsize (f)}   \\
\includegraphics[scale = 0.45, trim = 30 40 20 0]{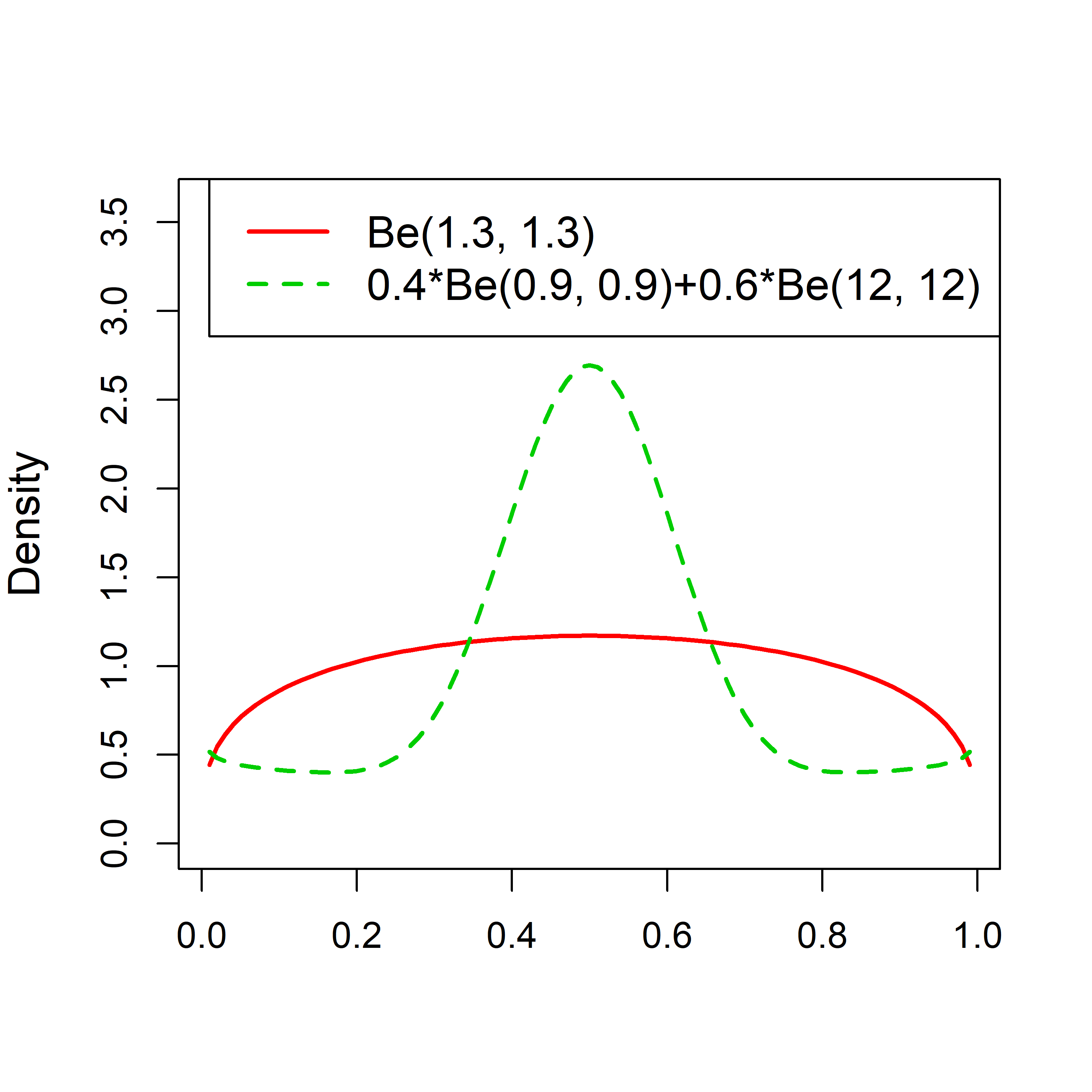} &
\includegraphics[scale = 0.45, trim = 30 40 20 0]{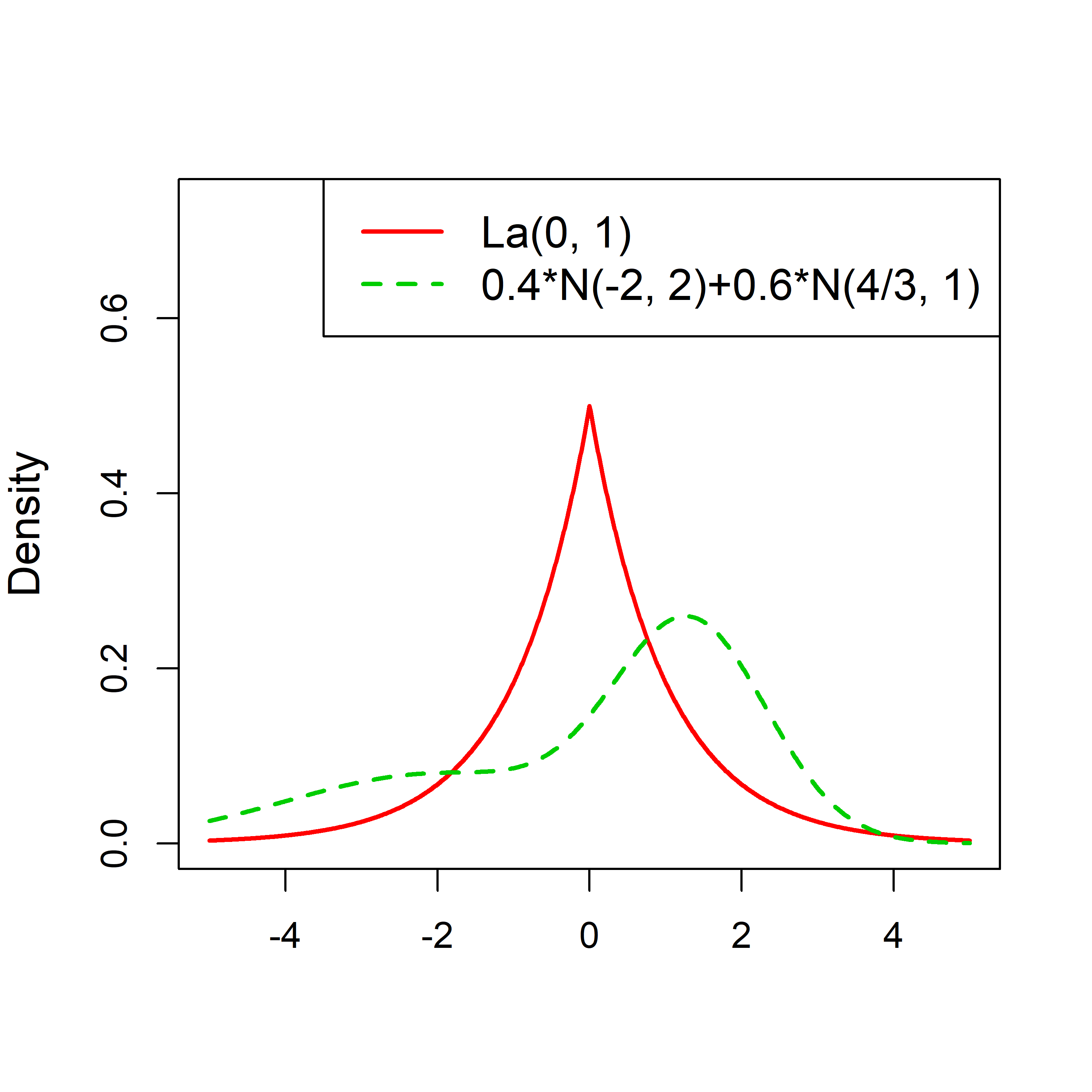} &
\includegraphics[scale = 0.45, trim = 40 40 20 0]{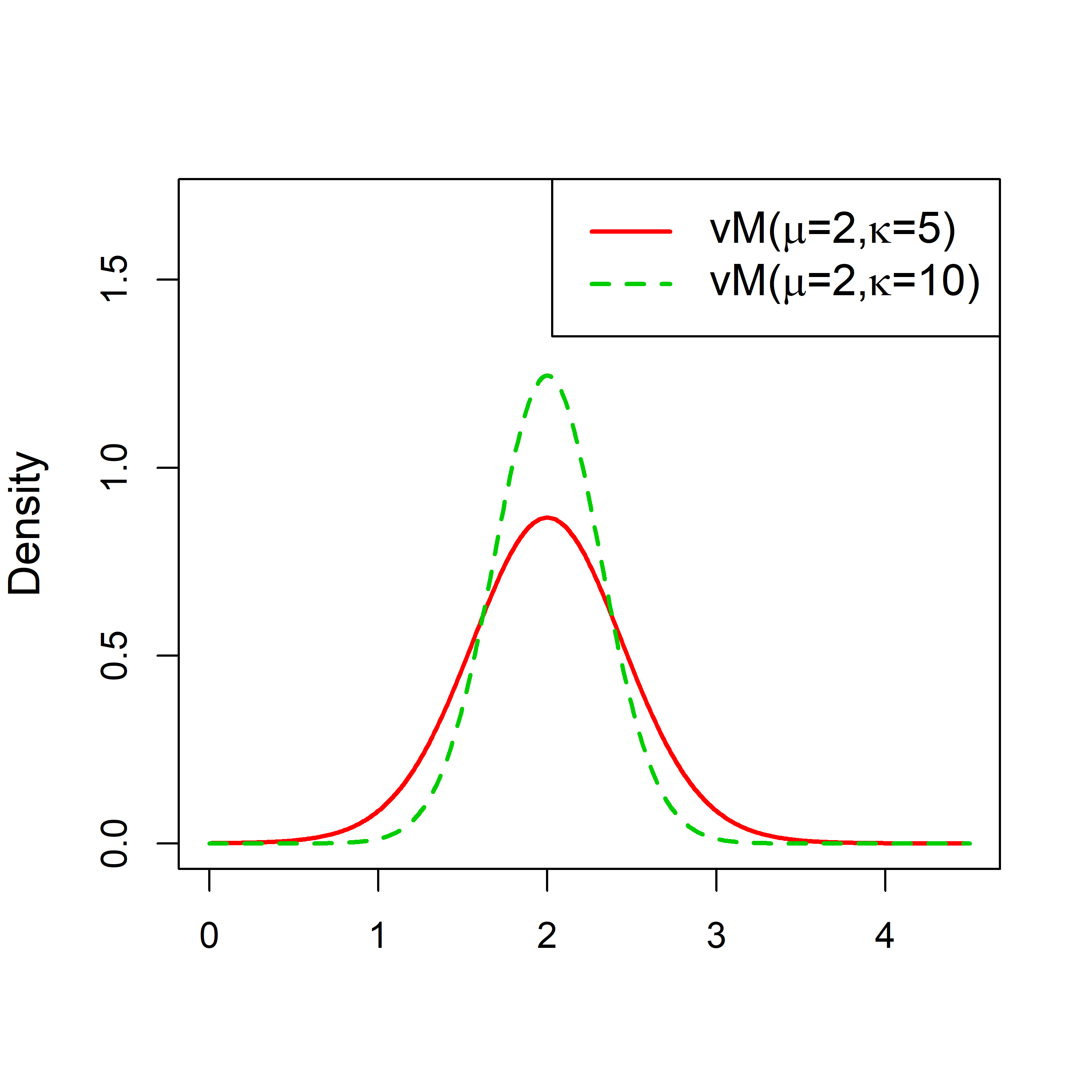}  \\
{\scriptsize (g)}   &   {\scriptsize (h)}  &   {\scriptsize (i)}   \\
\includegraphics[scale = 0.45, trim = 40 40 20 0]{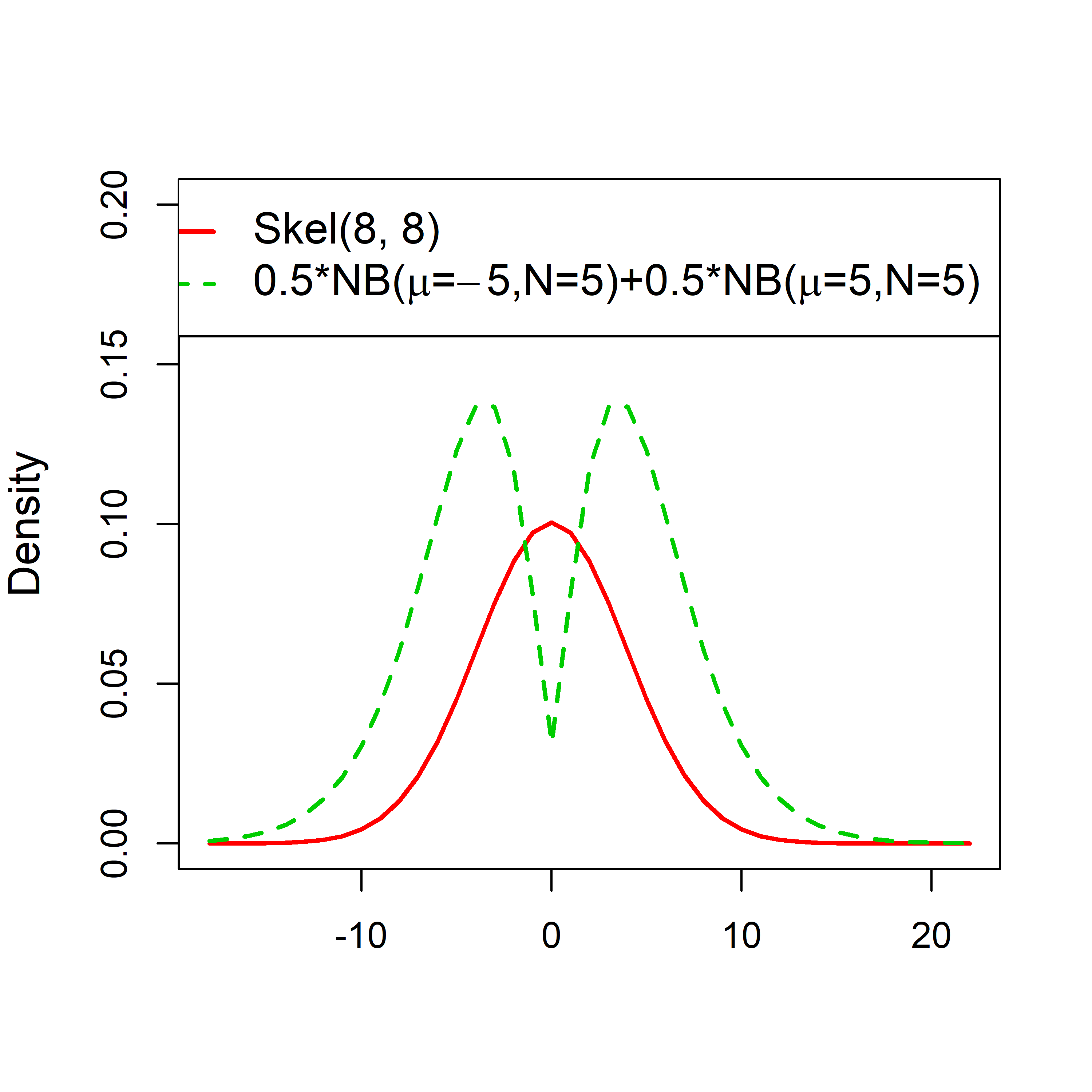} &
\includegraphics[scale = 0.45, trim = 30 40 20 0]{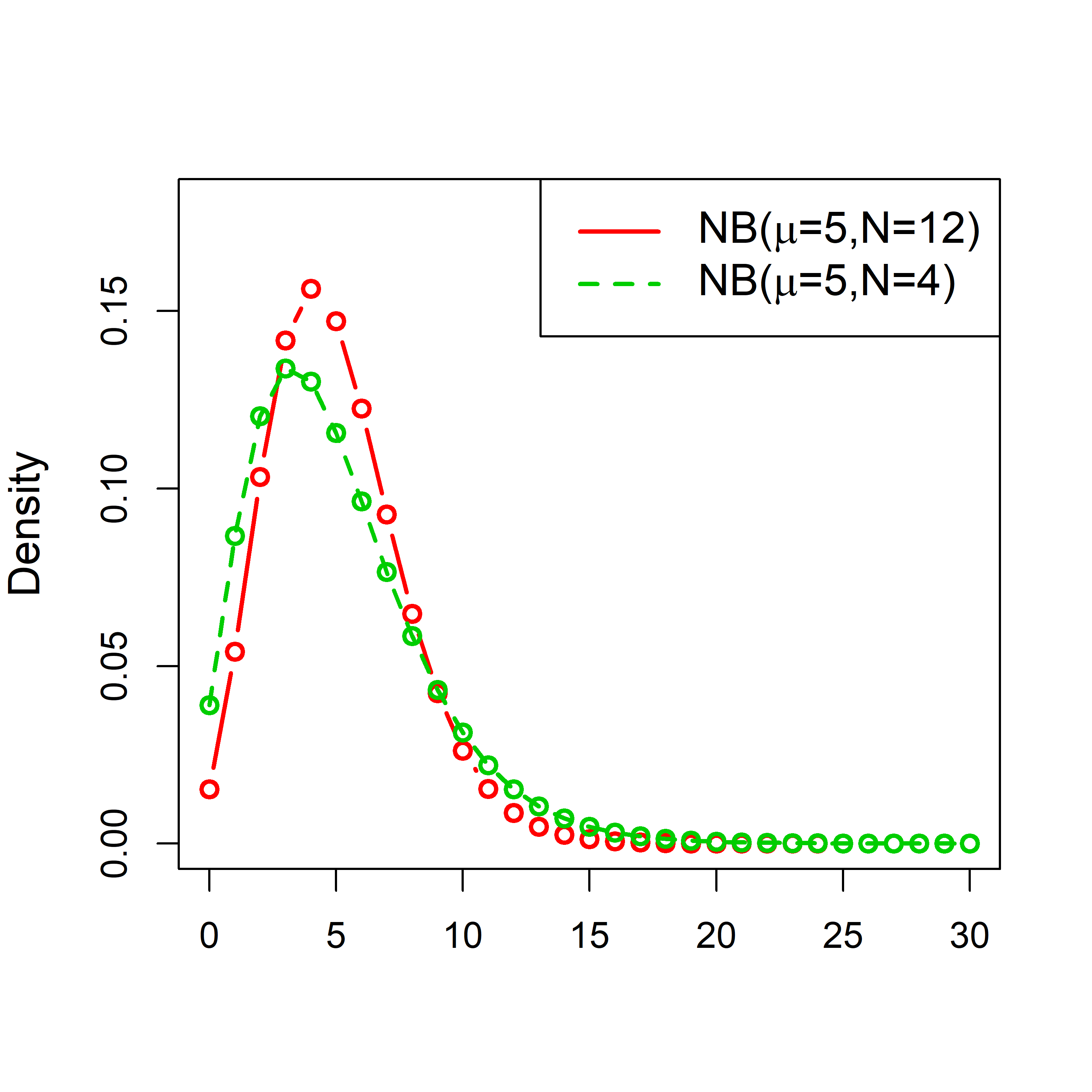} &
\includegraphics[scale = 0.45, trim = 30 40 20 0]{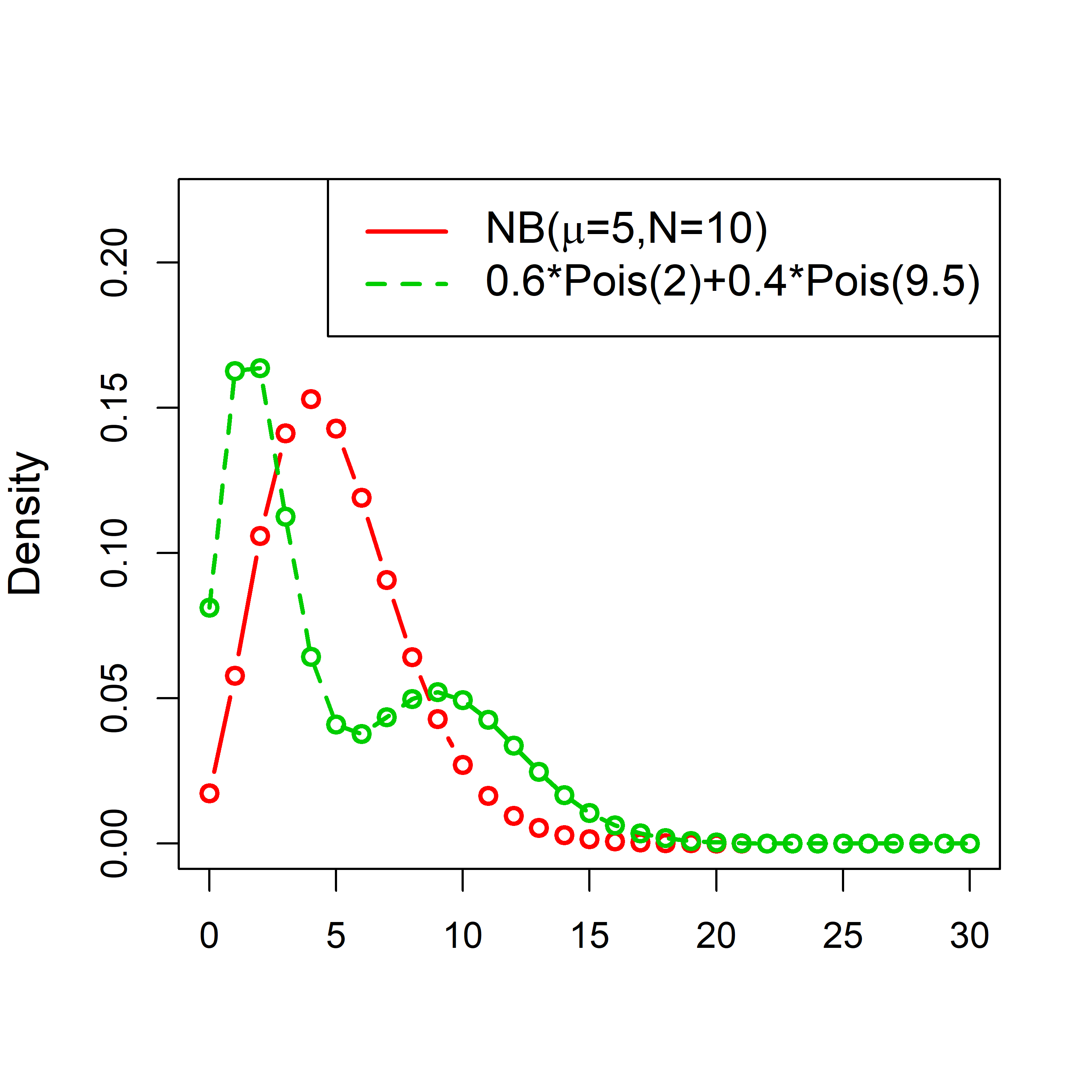}   \\
{\scriptsize (j)}   &   {\scriptsize (k)}  &   {\scriptsize (l)}   \\
\end{tabular}}
\caption{\textbf{The densities for each distributional scenario}.}
\label{fig.densities} 
\end{figure}

Twelve distributional scenarios are examined, all presented in Table \ref{tab.densities}, while Figure \ref{fig.densities} displays the relevant densities. The distributions, except for the first case scenario, are skewed and have high or low kurtosis values, while in some cases the distributions are bimodal. Further distributions with constrained support are considered, such as the beta, the gamma and the von Mises. The choice of these distributions was driven by the motivation of testing the procedures' robustness against deviations from normality. 

\begin{table}[htp]
\caption{\textbf{Distributional scenarios}. Each scenario is labeled by a letter that is appended on the first column. The second and third columns refer to the distributions used to generate each sample. Aiming at examining the robustness of the methods on distributions that vary from the normal, the normal distribution was used only in the first (a) scenario.}
\begin{center}
\begin{tabular}{lll}  \hline \hline
Scenario  &  Sample 1  &  Sample 2  \\   \hline
(a)  &  $N(3, 4)$  &  $N(3, 0.5)$  \\
(b)  &  $Be(3, 4)$  &  $N(3/7, 0.5)$  \\
(c)  &  $0.5Be(1, 5) + 0.5Be(5,2)$  &  $N(37/84, 0.5)$  \\
(d)  &  $Be(3, 4)$ & $0.5N(-11/7,0.5)+0.5N(17/7, 0.5)$  \\
(e)  &  $FN(3, 4)$  &  $Ga(4.049335, 1)$  \\ 
(f)  &  $0.5Ga(3, 3)+0.5Ga(8, 1)$ & $Ga(18, 4)$  \\ 
(g)  &  $Be(1.3, 1.3)$ & $0.4Be(0.9, 0.9)+0.6Be(12, 12)$  \\ 
(h)  &  $La(0, 1)$ & $0.4Be(-2, 2)+0.6N(/3, 1)$  \\ 
(i)  &  $vM(\mu=2, \kappa=10)$ & $vM(\mu=2, \kappa=5)$      \\
(j)  &  $Skel(\lambda_1=8,\lambda_2=8)$ & $0.5NB(\mu=-5, N=5)+0.5NB(\mu=5,N=5)$      \\
(k)  &  $NB(\mu=5, N=12)$ & $NB(\mu=5, N=4)$ \\ 
(l)  &  $NB(\mu=5, N=10)$ &  $0.6Pois(2) + 0.4Pois(9.5)$ \\ \hline \hline
\end{tabular}
\end{center}
\label{tab.densities}
\end{table}



\subsection{Type I error}
The estimated type I error for all cases is based on $1000$ repetitions and a test is termed size correct, when its type I error is correctly estimated, i.e. it falls within $\left(0.0365,0.0635\right)$\footnote{This is the asymptotically normal $95\%$ confidence interval for the true probability of type I error based on $1000$ simulations.}. Tables \ref{size} and \ref{size2} contain the estimated type I errors of all testing procedures with their calibrations and for all case scenarios.The WMW test almost always produced an inflated type I error, even when both populations are normal (Scenario (a)), and regardless of computing the asymptotic or the exact p-value. The EL and mainly the EEL, also suffered from inflated type I errors, whereas the Welch $t$-test was the only test that attained the type I error on most scenarios and sample sizes. 

Table \ref{size.perc} aggregates the estimated type I error for all scenarios. The WMW, regardless of the asymptotic or the exact p-value being computed, rarely attained the pre-specified type I error and EEL with the $\chi^2$ approximation was also inadequate. Both non parametric likelihoods benefited from the calibration with the $t$ distribution, with the EEL almost doubling the proportion of times it was size correct. EL with $t$ calibration and the Welch $t$-test were observed to be size correct in most cases. Finally, bootstrap calibration improved significantly the performance of the EL and EEL methods, while improving slightly the performance of the Welch $t$-test. 

\begin{figure}[htp]
\centering
\includegraphics[scale = 0.65, trim = 0 0 0 0]{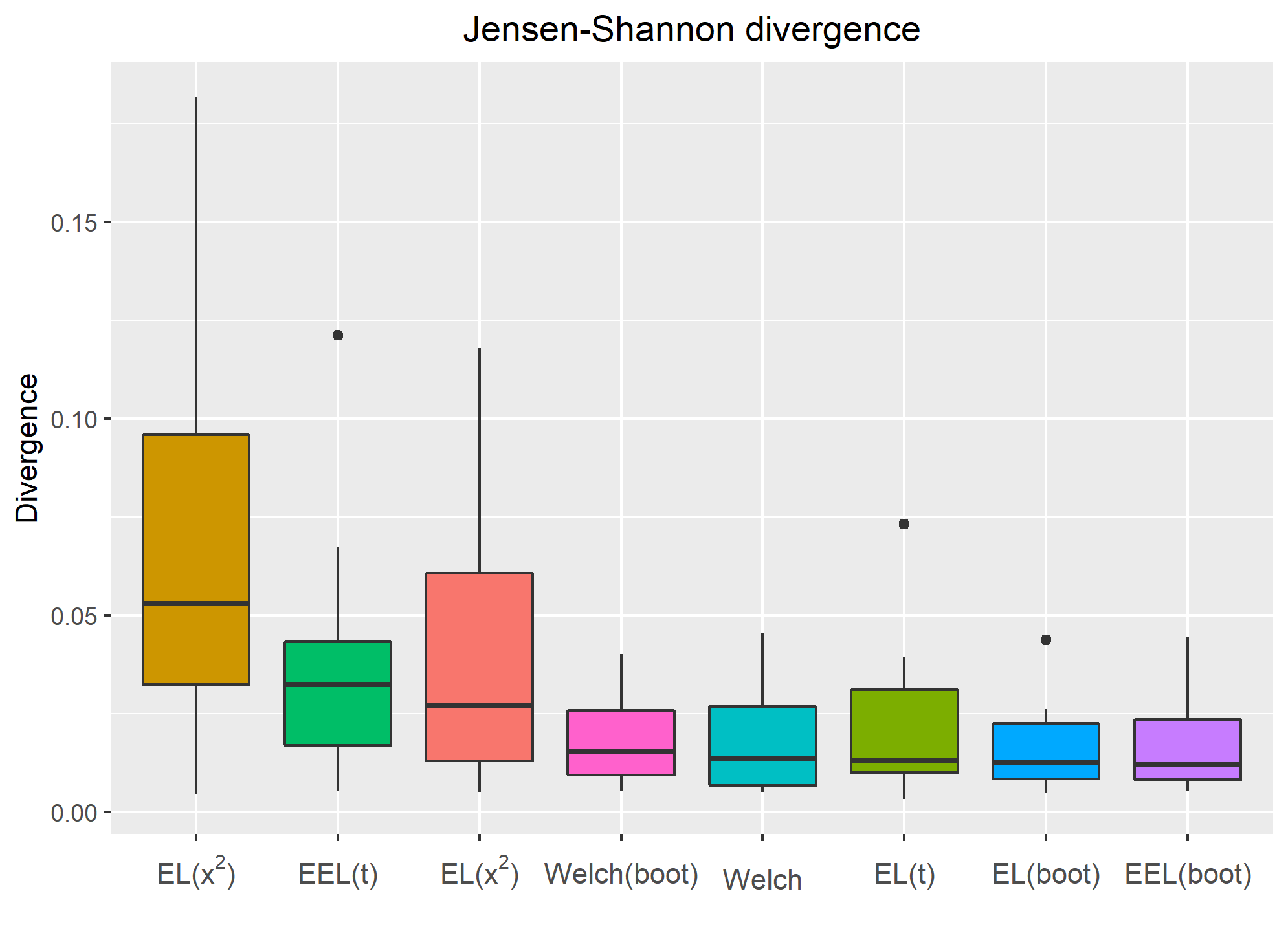}  \\
\includegraphics[scale = 0.65, trim = 0 0 0 0]{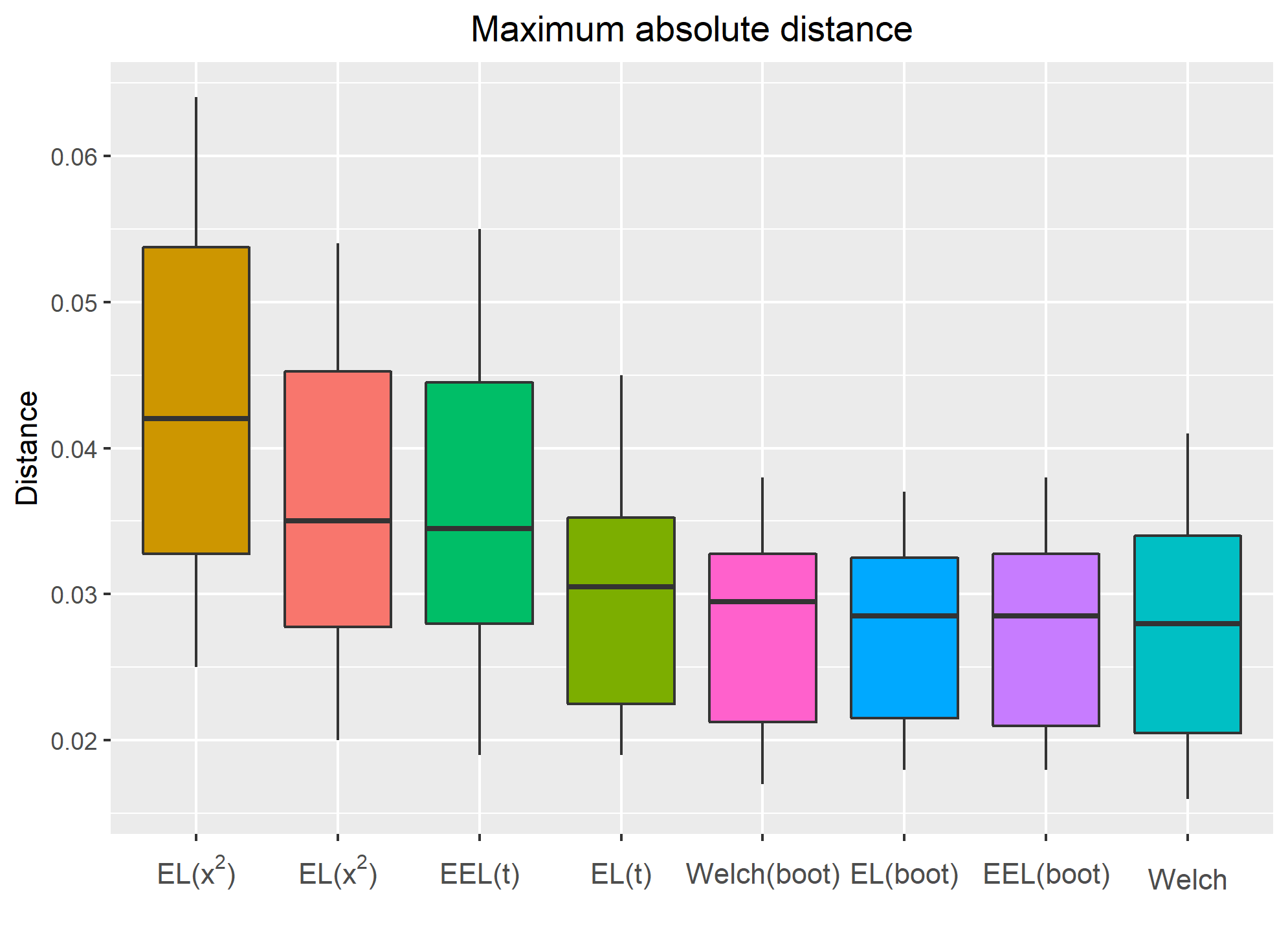}  
\caption{\textbf{Jensen-Shannon divergence (up) and maximum absolute distance (down) between the true and the observed quantiles of the distribution of the test statistics.}}
\label{klmd} 
\end{figure}

It is not uncommon for a test to accurately estimate the 5\% quantile of the limiting distribution correctly, but not the 10\% or 1\% quantile. The EL, EEL and WMW test even though overestimated the type I error at the 5\% level, they were estimating correctly the upper tail of the distribution. To address this point, the type I error was estimated at numerous levels, from 1\% up to 99\%. Figure \ref{klmd} presents the Jensen-Shannon divergence and the maximum absolute difference between the true quantiles and the estimated quantiles of each test statistic. The WMW (using the asymptotic and the exact p-value) was excluded because it was shown to be very inaccurate. The EL calibrated with the $t$ distribution, the Welch $t$-test and the EL, EEL and Welch $t$-test with bootstrap calibration were the test statistics whose estimated quantiles were highly accurate. This information strengthens the results of the simulation studies, since these test statistics accurately estimate the null distribution and not only its left tail. 

\setcounter{table}{1}
\begin{sidewaystable}[htp]
\caption{\textbf{Estimated probability of Type I error using different tests and a variety of calibrations for the first 6 scenarios}. Each row refers to scenarios (a)-(f) described in Table \ref{tab.densities}, where for each scenario, 5 different sample sizes pairs were examined. The first column refers to the size of each sample. The rest of the columns refer to the estimated probability of type I error (nominal level is $0.05$) of each testing procedure applied on each scenario. The numbers in bold indicate that the estimated probability was within the acceptable limits $\left(0.0365,0.0635\right)$. The calibration technique, if one was applied is mentioned inside the parentheses. The $\chi^2$ refers to the $\chi^2$ distribution, $t$ refers to the $t$ distribution, "boot" refers to bootstrap calibration and "exact" refers to the exact calculation of the p-value of the WMW test.}
{\begin{tabular}{@{}l|l|lccccccccc}  \\ \hline \hline
& Sample  & \multicolumn{10}{c}{\underline{Testing procedures}} \\ 
& sizes   & EL($\chi^2$) & EEL($\chi^2$) & EL($t$) & EEL($t$) &  Welch & WMW & EL(boot) & EEL(boot) & Welch(boot)  & WMW(exact) \\ \hline
\multirow{5}{*}{Scenario (a)}  
& (20, 30)  & 0.065 & 0.076 & \textbf{0.050} & \textbf{0.061} & \textbf{0.048} & 0.120 & \textbf{0.046} & \textbf{0.046} & \textbf{0.047} & 0.120  \\
& (20, 40)  & 0.081 & 0.085 & 0.065 & 0.073 & 0.065 & 0.138 & 0.064 & \textbf{0.062} & \textbf{0.056} & 0.138  \\
& (30, 40)  & \textbf{0.060} & 0.069 & \textbf{0.049} & \textbf{0.057} & \textbf{0.052} & 0.121 & \textbf{0.048} & \textbf{0.048} & \textbf{0.049} & 0.121  \\
& (30, 50)  & \textbf{0.048} & \textbf{0.052} & \textbf{0.039} & \textbf{0.043} & \textbf{0.039} & 0.122 & \textbf{0.042} & \textbf{0.044} & \textbf{0.038} & 0.122 \\
& (50, 100) & \textbf{0.057} & \textbf{0.062} & \textbf{0.055} & \textbf{0.056} & \textbf{0.052} & 0.149 & \textbf{0.052} & \textbf{0.052} & \textbf{0.051} & 0.150 \\ \hline 
\multirow{5}{*}{Scenario (b)}   
& (20, 30)  & \textbf{0.058} & 0.066 & \textbf{0.053} & \textbf{0.057} & \textbf{0.051} & 0.033 & \textbf{0.052} & \textbf{0.052} & \textbf{0.048} & 0.033 \\
& (20, 40)  & \textbf{0.043} & \textbf{0.049} & \textbf{0.039} & \textbf{0.047} & \textbf{0.038} & 0.024 & \textbf{0.038} & 0.034 & \textbf{0.038} & 0.024 \\
& (30, 40)  & \textbf{0.044} & \textbf{0.048} & \textbf{0.042} & \textbf{0.045} & \textbf{0.043} & \textbf{0.050} & \textbf{0.037} & \textbf{0.037} & \textbf{0.040} & \textbf{0.050} \\
& (30  50)  & \textbf{0.055} & \textbf{0.062} & \textbf{0.051} & \textbf{0.056} & \textbf{0.052} & \textbf{0.038} & \textbf{0.048} & \textbf{0.047} & \textbf{0.051} & \textbf{0.038} \\
& (50, 100) & \textbf{0.060} & 0.069 & \textbf{0.060} & 0.067 & \textbf{0.058} & 0.029 & \textbf{0.060} & \textbf{0.060} & \textbf{0.057} & 0.029 \\ \hline
\multirow{5}{*}{Scenario (c)}   
& (20, 30)    &  0.076 & 0.083 & 0.066 & 0.074 & \textbf{0.062} & 0.071 & 0.065 & 0.065 & \textbf{0.055} & 0.071 \\
& (20, 40)   &  \textbf{0.062} & 0.083 & \textbf{0.053} & 0.073 & \textbf{0.048} & 0.068 & \textbf{0.052} & \textbf{0.051} & \textbf{0.051} & 0.068 \\
& (30, 40)   &  \textbf{0.053} & \textbf{0.058} & \textbf{0.047} & \textbf{0.054} & \textbf{0.047} & \textbf{0.052} & \textbf{0.043} & \textbf{0.044} & \textbf{0.044} & \textbf{0.052} \\
& (30, 50)   &  0.065 & 0.073 & \textbf{0.060} & 0.067 & \textbf{0.058} & 0.072 & \textbf{0.060} & \textbf{0.059} & \textbf{0.056} & 0.072 \\
& (50, 100)  &  \textbf{0.043} & 0.066 & \textbf{0.040} & \textbf{0.063} & \textbf{0.041} & \textbf{0.063} & \textbf{0.038} & \textbf{0.038} & \textbf{0.043} & \textbf{0.063} \\ \hline
\multirow{5}{*}{Scenario (d)}   
& (20, 30)  & \textbf{0.050} & \textbf{0.052} & \textbf{0.037} & \textbf{0.037} & \textbf{0.049} & 0.102 & \textbf{0.041} & \textbf{0.041} & \textbf{0.033} & 0.102 \\
& (20, 40)  & \textbf{0.055} & \textbf{0.057} & \textbf{0.039} & \textbf{0.044} & \textbf{0.052} & 0.113 & \textbf{0.048} & \textbf{0.047} & \textbf{0.039} & 0.113 \\
& (30, 40)  & \textbf{0.054} & \textbf{0.054} & \textbf{0.042} & \textbf{0.042} & \textbf{0.054} & 0.108 & \textbf{0.050} & \textbf{0.050} & \textbf{0.045} & 0.107 \\ 
& (30, 50)  & \textbf{0.045} & \textbf{0.047} & 0.031 & 0.034 & \textbf{0.042} & 0.195 & \textbf{0.044} & \textbf{0.042} & \textbf{0.036} & 0.195 \\
& (50, 100) & \textbf{0.050} & \textbf{0.050} & \textbf{0.042} & \textbf{0.045} & \textbf{0.048} & 0.195 & \textbf{0.042} & \textbf{0.041} & \textbf{0.037} & 0.195 \\ \hline
\multirow{5}{*}{Scenario (e)}  
& (20, 30)  & \textbf{0.058} & 0.074 & \textbf{0.048} & \textbf{0.062} & \textbf{0.043} & 0.079 & \textbf{0.043} & \textbf{0.041} & \textbf{0.039} & 0.079 \\
& (20, 40)  & \textbf{0.055} & 0.076 & \textbf{0.045} & \textbf{0.060} & \textbf{0.047} & 0.098 & \textbf{0.043} & \textbf{0.042} & \textbf{0.040} & 0.098 \\
& (30, 40)  & \textbf{0.051} & \textbf{0.057} & \textbf{0.047} & \textbf{0.048} & \textbf{0.046} & 0.086 & \textbf{0.047} & \textbf{0.047} & \textbf{0.045} & 0.086 \\
& (30, 50)  & \textbf{0.039} & \textbf{0.045} & 0.034 & \textbf{0.041} & 0.032 & 0.084 & 0.030 & 0.030 & 0.028 & 0.084 \\
& (50, 100) & \textbf{0.041} & \textbf{0.057} & \textbf{0.038} & \textbf{0.055} & \textbf{0.041} & 0.124 & \textbf{0.038} & \textbf{0.039} & \textbf{0.042} & 0.125 \\ \hline 
\multirow{5}{*}{Scenario (f)}   
& (20, 30)  & 0.070 & 0.074 & \textbf{0.047} & \textbf{0.054} & \textbf{0.058} & 0.161 & \textbf{0.049} & \textbf{0.049} & \textbf{0.042} & 0.161 \\
& (20, 40)  & 0.071 & 0.081 & \textbf{0.055} & 0.064 & 0.066 & 0.195 & \textbf{0.053} & \textbf{0.051} & \textbf{0.049} & 0.195 \\
& (30, 40)  & \textbf{0.055} & \textbf{0.059} & \textbf{0.050} & \textbf{0.051} & \textbf{0.055} & 0.159 & \textbf{0.051} & \textbf{0.050} & \textbf{0.050} & 0.159 \\
& (30, 50)  & \textbf{0.053} & \textbf{0.059} & \textbf{0.044} & \textbf{0.049} & \textbf{0.048} & 0.197 & \textbf{0.043} & \textbf{0.043} & \textbf{0.044} & 0.197 \\
& (50, 100) & 0.067 & 0.076 & \textbf{0.063} & 0.068 & 0.072 & 0.275 & 0.067 & 0.069 & 0.066 & 0.276 \\ \hline \hline
\end{tabular}}
\label{size} 
\end{sidewaystable}

\setcounter{table}{2}
\begin{sidewaystable}
\caption{\textbf{Estimated probability of Type I error using different tests and a variety of calibrations for the last 6 scenarios}. Each row refers to scenarios (g)-(l) described in Table \ref{tab.densities}, where for each scenario, 5 different sample sizes pairs were examined. The first column refers to the size of each sample. The rest of the columns refer to the estimated probability of type I error (nominal level is $0.05$) of each testing procedure applied on each scenario. The numbers in bold indicate that the estimated probability was within the acceptable limits $\left(0.0365,0.0635\right)$. The calibration technique, if one was applied is mentioned inside the parentheses. The $\chi^2$ refers to the $\chi^2$ distribution, $t$ refers to the $t$ distribution, "boot" refers to bootstrap calibration and "exact" refers to the exact calculation of the p-value of the WMW test. The "-" indicates that the exact p-value for the WMW test statistic was not computed because of ties in the data.}
{\begin{tabular}{l|l|cccccccccc}  \\  \hline \hline
& Sample  & \multicolumn{10}{c}{\underline{Testing procedures}} \\ 
& sizes   & EL($\chi^2$) & EEL($\chi^2$) & EL($t$) & EEL($t$) &  Welch & WMW & EL(boot) & EEL(boot) & Welch(boot)  & WMW(exact) \\  \hline
\multirow{5}{*}{Scenario (g)}  
& (20, 30)  & \textbf{0.055} & 0.064 & \textbf{0.046} & \textbf{0.054} & \textbf{0.046} & \textbf{0.062} & \textbf{0.045} & \textbf{0.046} & \textbf{0.044} & \textbf{0.062} \\
& (20, 40)  & 0.069 & 0.088 & \textbf{0.055} & 0.075 & \textbf{0.061} & 0.085 & \textbf{0.054} & \textbf{0.057} & \textbf{0.055} & 0.085 \\
& (30, 40)  & \textbf{0.047} & \textbf{0.054} & \textbf{0.044} & \textbf{0.047} & \textbf{0.045} & \textbf{0.059} & \textbf{0.046} & \textbf{0.047} & \textbf{0.047} & 0.059 \\
& (30, 50)  & \textbf{0.059} & 0.069 & \textbf{0.052} & \textbf{0.062} & \textbf{0.057} & 0.074 & \textbf{0.051} & \textbf{0.051} & \textbf{0.052} & 0.074 \\
& (50, 100) & 0.069 & 0.085 & 0.064 & 0.082 & 0.067 & 0.085 & \textbf{0.059} & \textbf{0.060} & \textbf{0.058} & 0.086 \\ \hline 
\multirow{5}{*}{Scenario (h)}  
& (20, 30)  & \textbf{0.062} & 0.071 & \textbf{0.054} & 0.064 & \textbf{0.047} & 0.119 & \textbf{0.047} & \textbf{0.048} & \textbf{0.045} & 0.119 \\
& (20, 40)  & \textbf{0.060} & 0.084 & \textbf{0.048} & 0.069 & \textbf{0.052} & 0.153 & \textbf{0.048} & \textbf{0.046} & \textbf{0.045} & 0.153 \\
& (30, 40)  & \textbf{0.058} & 0.066 & \textbf{0.051} & \textbf{0.058} & \textbf{0.056} & 0.174 & \textbf{0.049} & \textbf{0.052} & \textbf{0.054} & 0.174 \\
& (30, 50)  & \textbf{0.051} & \textbf{0.062} & \textbf{0.038} & \textbf{0.054} & \textbf{0.045} & 0.185 & \textbf{0.041} & \textbf{0.041} & \textbf{0.040} & 0.185 \\
& (50, 100) & \textbf{0.056} & 0.077 & \textbf{0.052} & 0.070 & \textbf{0.055} & 0.287 & \textbf{0.051} & \textbf{0.050} & \textbf{0.048} & 0.289 \\ \hline
\multirow{5}{*}{Scenario (i)}  
& (20, 30)  & \textbf{0.061} & 0.073 & \textbf{0.048} & \textbf{0.059} & \textbf{0.048} & \textbf{0.059} & \textbf{0.048} & \textbf{0.048} & \textbf{0.045} & \textbf{0.059} \\
& (20, 40)  & \textbf{0.052} & 0.071 & \textbf{0.043} & \textbf{0.056} & \textbf{0.037} & \textbf{0.055} & \textbf{0.039} & \textbf{0.040} & \textbf{0.034} & \textbf{0.055} \\
& (30, 40)  & \textbf{0.062} & 0.067 & \textbf{0.059} & \textbf{0.062} & \textbf{0.057} & \textbf{0.056} & \textbf{0.055} & \textbf{0.055} & \textbf{0.052} & \textbf{0.056} \\
& (30, 50)  & \textbf{0.058} & 0.067 & \textbf{0.055} & \textbf{0.059} & \textbf{0.052} & 0.067 & \textbf{0.042} & \textbf{0.044} & \textbf{0.046} & 0.067 \\
& (50, 100) & \textbf{0.058} & 0.079 & \textbf{0.050} & 0.073 & \textbf{0.052} & 0.069 & \textbf{0.050} & \textbf{0.048} & \textbf{0.047} & 0.069 \\ \hline
\multirow{5}{*}{Scenario (j)}  
& (20, 30)  & \textbf{0.057} & \textbf{0.062} & \textbf{0.049} & \textbf{0.058} & \textbf{0.051} & \textbf{0.059} & \textbf{0.043} & \textbf{0.042} & \textbf{0.044} & - \\
& (20, 40)  & \textbf{0.059} & 0.080 & \textbf{0.049} & 0.069 & \textbf{0.053} & 0.071 & \textbf{0.049} & \textbf{0.046} & \textbf{0.045} & - \\
& (30, 40)  & \textbf{0.059} & \textbf{0.064} & \textbf{0.054} & \textbf{0.058} & \textbf{0.053} & 0.066 & \textbf{0.050} & \textbf{0.050} & \textbf{0.051} & - \\
& (30, 50)  & \textbf{0.050} & \textbf{0.062} & \textbf{0.045} & \textbf{0.053} & \textbf{0.045} & \textbf{0.063} & \textbf{0.043} & \textbf{0.042} & \textbf{0.041} & - \\
& (50, 100) & \textbf{0.061} & 0.073 & \textbf{0.055} & 0.070 & \textbf{0.058} & 0.082 & \textbf{0.053} & \textbf{0.053} & \textbf{0.054} & - \\ \hline
\multirow{5}{*}{Scenario (k)}  
& (20, 30)  & 0.066 & 0.074 & \textbf{0.059} & 0.065 & \textbf{0.060} & 0.066 & \textbf{0.058} & \textbf{0.059} & \textbf{0.056} & - \\
& (20, 40)  & \textbf{0.052} & 0.065 & \textbf{0.042} & \textbf{0.063} & \textbf{0.046} & \textbf{0.063} & \textbf{0.042} & \textbf{0.043} & \textbf{0.039} & - \\
& (30, 40)  & \textbf{0.058} & 0.066 & \textbf{0.051} & \textbf{0.057} & \textbf{0.051} & 0.067 & \textbf{0.050} & \textbf{0.054} & \textbf{0.051} & - \\
& (30, 50)  & \textbf{0.046} & \textbf{0.060} & \textbf{0.042} & \textbf{0.056} & \textbf{0.045} & \textbf{0.056} & \textbf{0.040} & \textbf{0.039} & 0.032 & - \\
& (50, 100) & \textbf{0.059} & 0.073 & \textbf{0.056} & 0.068 & \textbf{0.054} & 0.080 & \textbf{0.056} & \textbf{0.055} & \textbf{0.050} & - \\ \hline
\multirow{5}{*}{Scenario (l)}  
& (20, 30)  & 0.073 & 0.079 & \textbf{0.061} & 0.072 & 0.067 & 0.151 & \textbf{0.058} & \textbf{0.054} & \textbf{0.056} & - \\
& (20, 40)  & \textbf{0.049} & 0.075 & \textbf{0.038} & \textbf{0.062} & \textbf{0.056} & 0.137 & \textbf{0.039} & \textbf{0.037} & \textbf{0.042} & - \\
& (30, 40)  & \textbf{0.056} & \textbf{0.061} & \textbf{0.048} & \textbf{0.053} & \textbf{0.049} & 0.134 & \textbf{0.044} & \textbf{0.046} & \textbf{0.046} & - \\
& (30, 50)  & \textbf{0.057} & 0.069 & \textbf{0.046} & \textbf{0.058} & \textbf{0.049} & 0.160 & \textbf{0.049} & \textbf{0.044} & \textbf{0.045} & - \\
& (50, 100) & \textbf{0.057} & 0.073 & \textbf{0.052} & 0.067 & \textbf{0.056} & 0.263 & \textbf{0.058} & \textbf{0.059} & \textbf{0.057} & - \\ \hline \hline
\end{tabular}}
\label{size2} 
\end{sidewaystable}

\begin{table}[htp]
\caption{\textbf{Number of times each testing procedure attained the correct size (type I error)}. The first and third rows refer to each testing procedure applied on the scenario. The $\chi^2$ and $t$ refer to calibration with the $\chi^2$ and the $t$ distributions respectively, "boot" refers to bootstrap calibration and "exact" refers to the exact calculation of the p-value of the WMW test. The second and fourth rows show the proportion of times a testing procedure was size correct. The value 60 in the denominator comes from applying 5 different sample sizes on 12 scenarios.}
\begin{center}
{\begin{tabular}{l|ccccc} \hline \hline
Testing procedure  & EL$(\chi^2)$ & EEL$(\chi^2)$  & EL$(t)$   & EEL$(t)$    & Welch  \\  \hline                                     Proportion         &  49/60       &  23/60         &  55/60    & 39/60       & 54/60  \\  \hline \hline
Testing procedure  & WMW          & EL(boot)       & EEL(boot) & Welch(boot) & WMW(exact)  \\  
Proportion         & 13/60        & 56/60          & 56/60     & 57/60       &  8/45      \\ \hline \hline
\end{tabular}}
\end{center}
\label{size.perc}
\end{table}

\subsection{Power of the Welch $t$-test, EL and EEL with bootstrap calibration}
\begin{figure}[htp]
\centering
{\begin{tabular}{ccc}
\includegraphics[scale = 0.45, trim = 40 30 20 0]{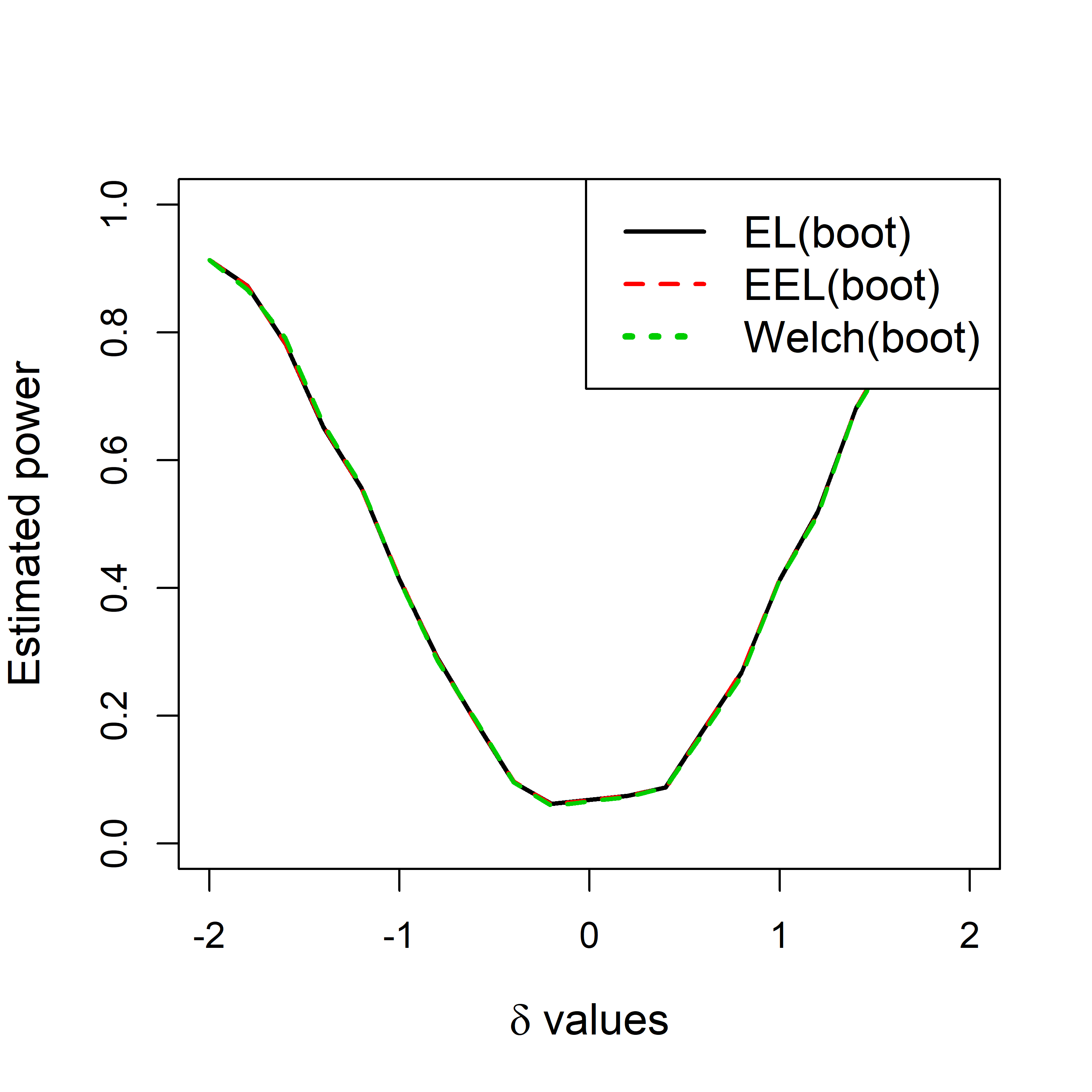} &
\includegraphics[scale = 0.45, trim = 30 30 20 0]{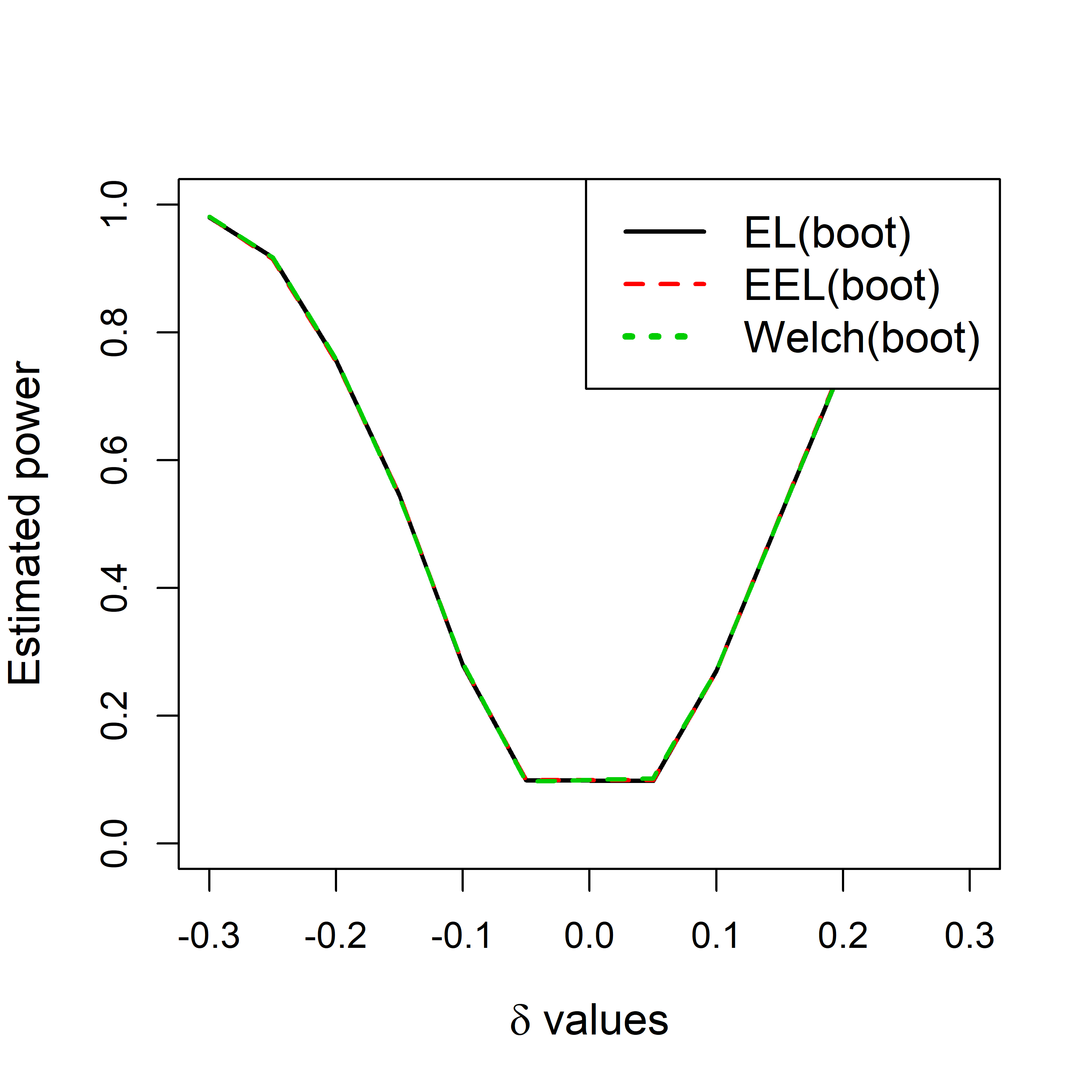} &
\includegraphics[scale = 0.45, trim = 30 30 0 0]{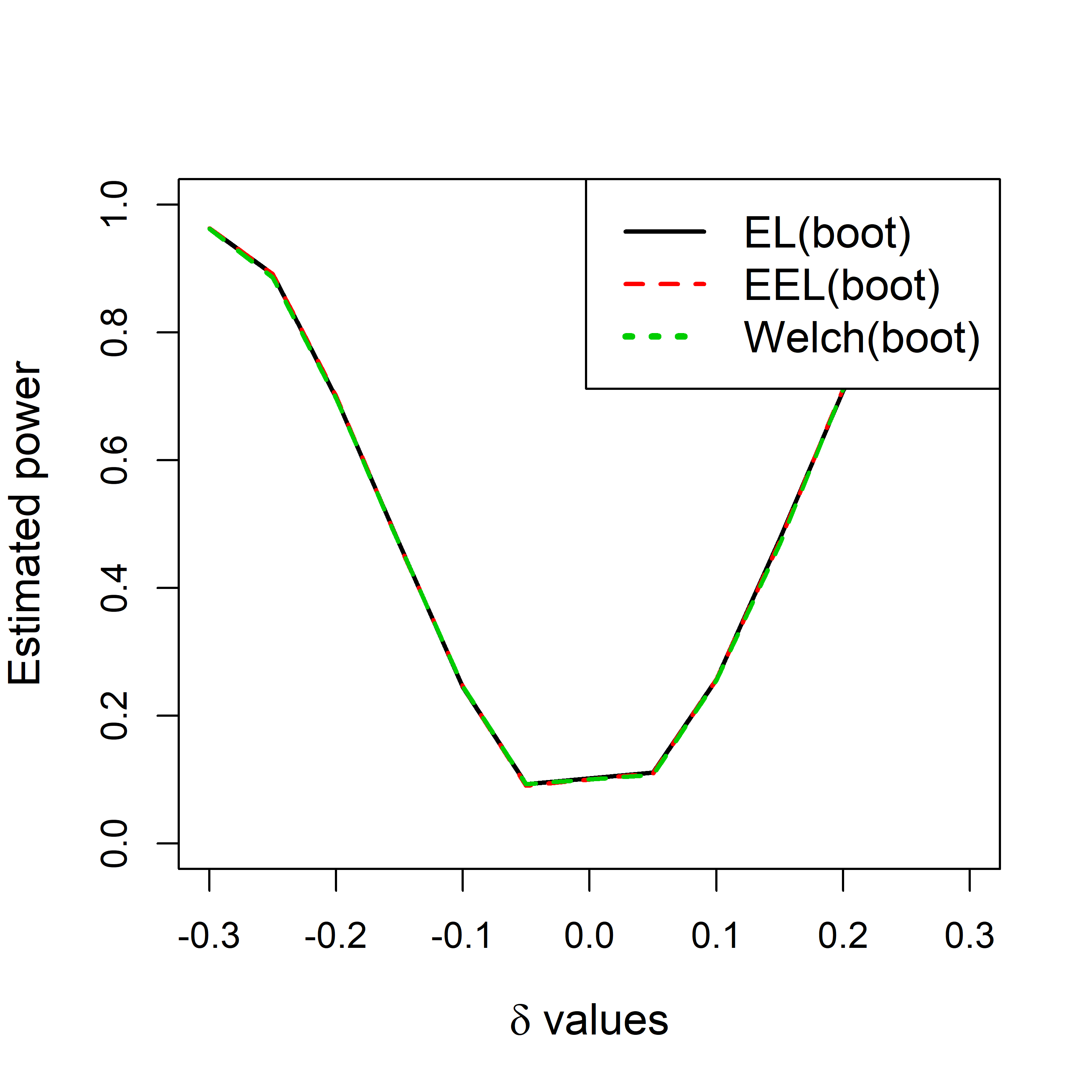}   \\
{\scriptsize (a)}   &   {\scriptsize (b)}  &   {\scriptsize (c)}   \\
\includegraphics[scale = 0.45, trim = 40 30 20 5]{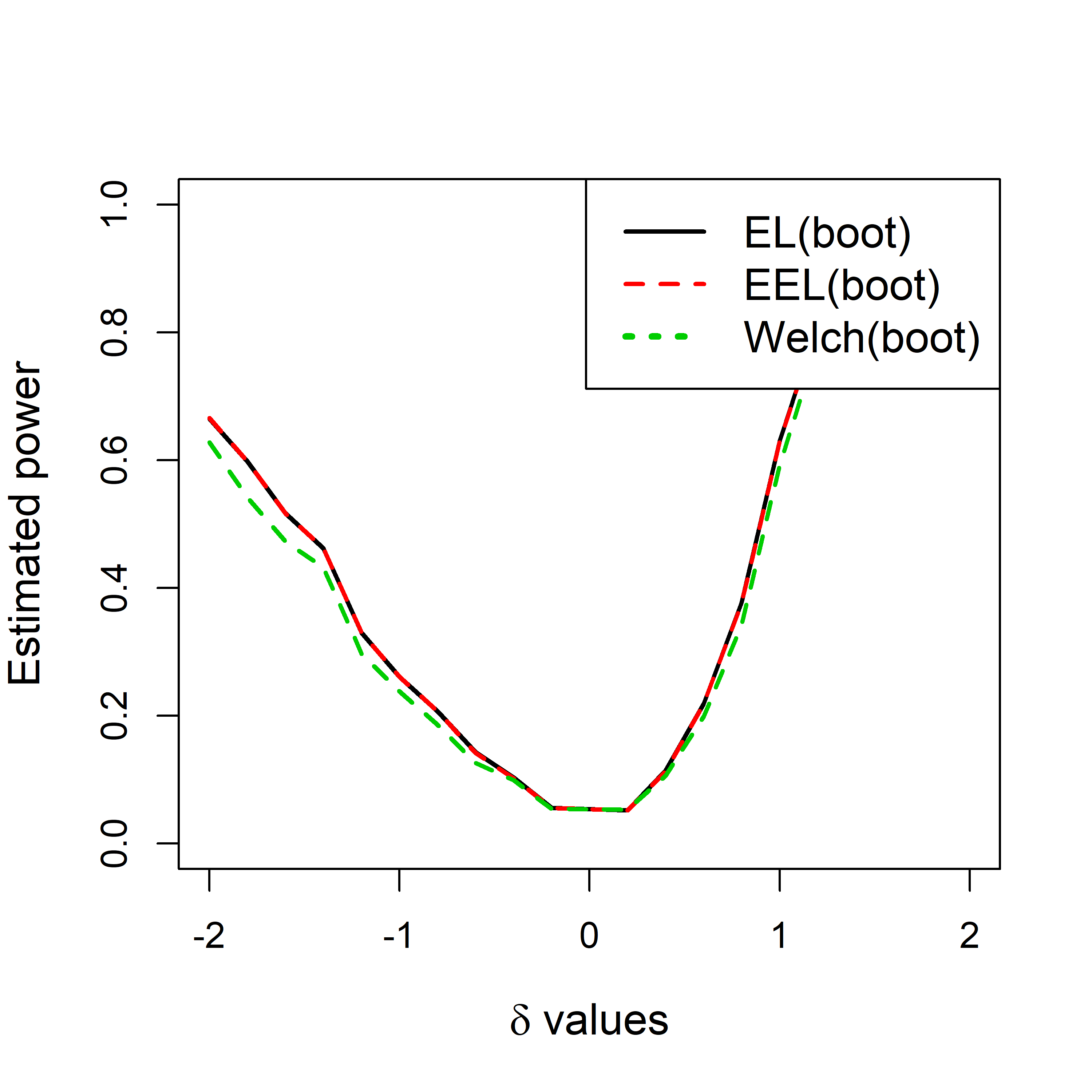} &  
\includegraphics[scale = 0.45, trim = 30 30 0 5]{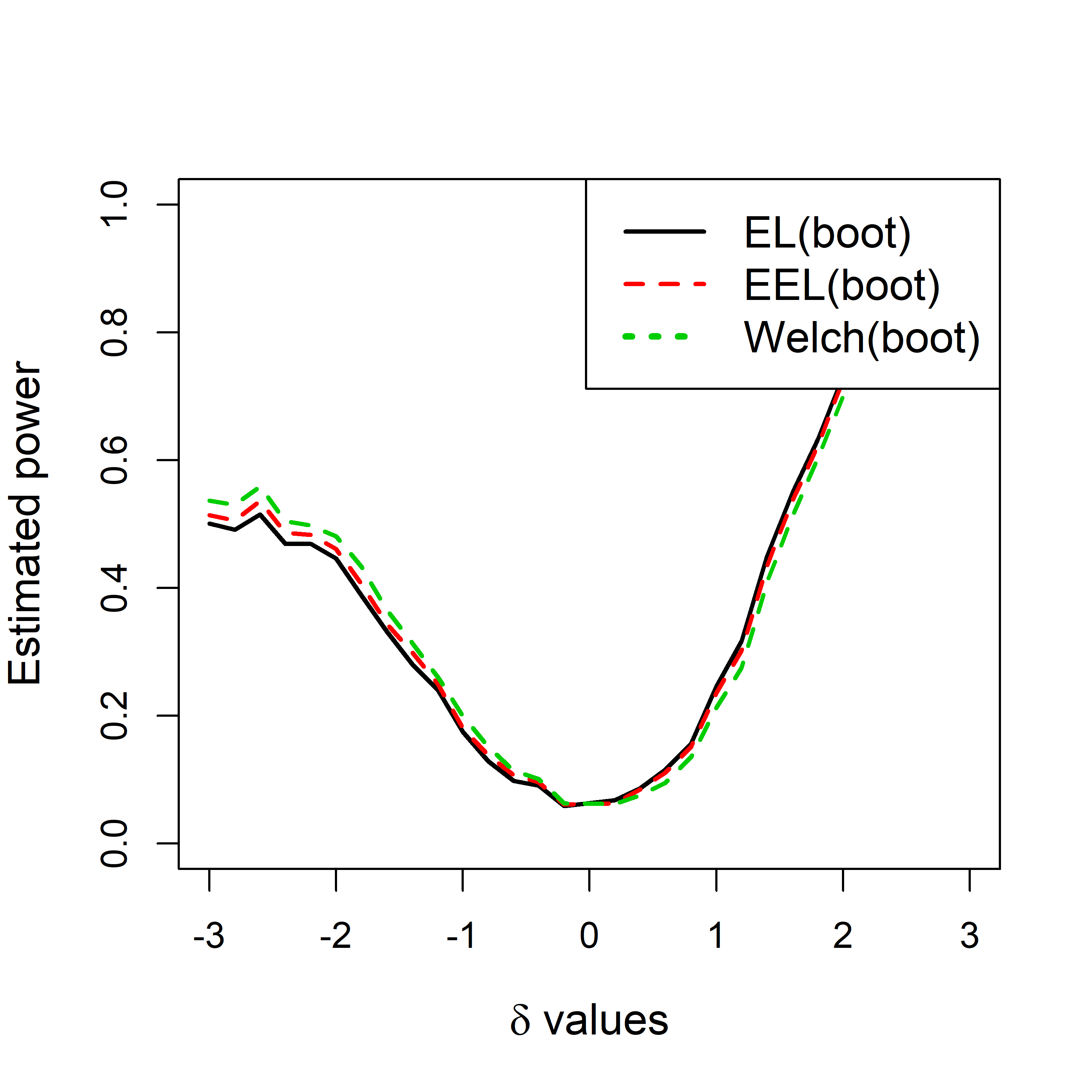}  &  
\includegraphics[scale = 0.45, trim = 30 30 0 5]{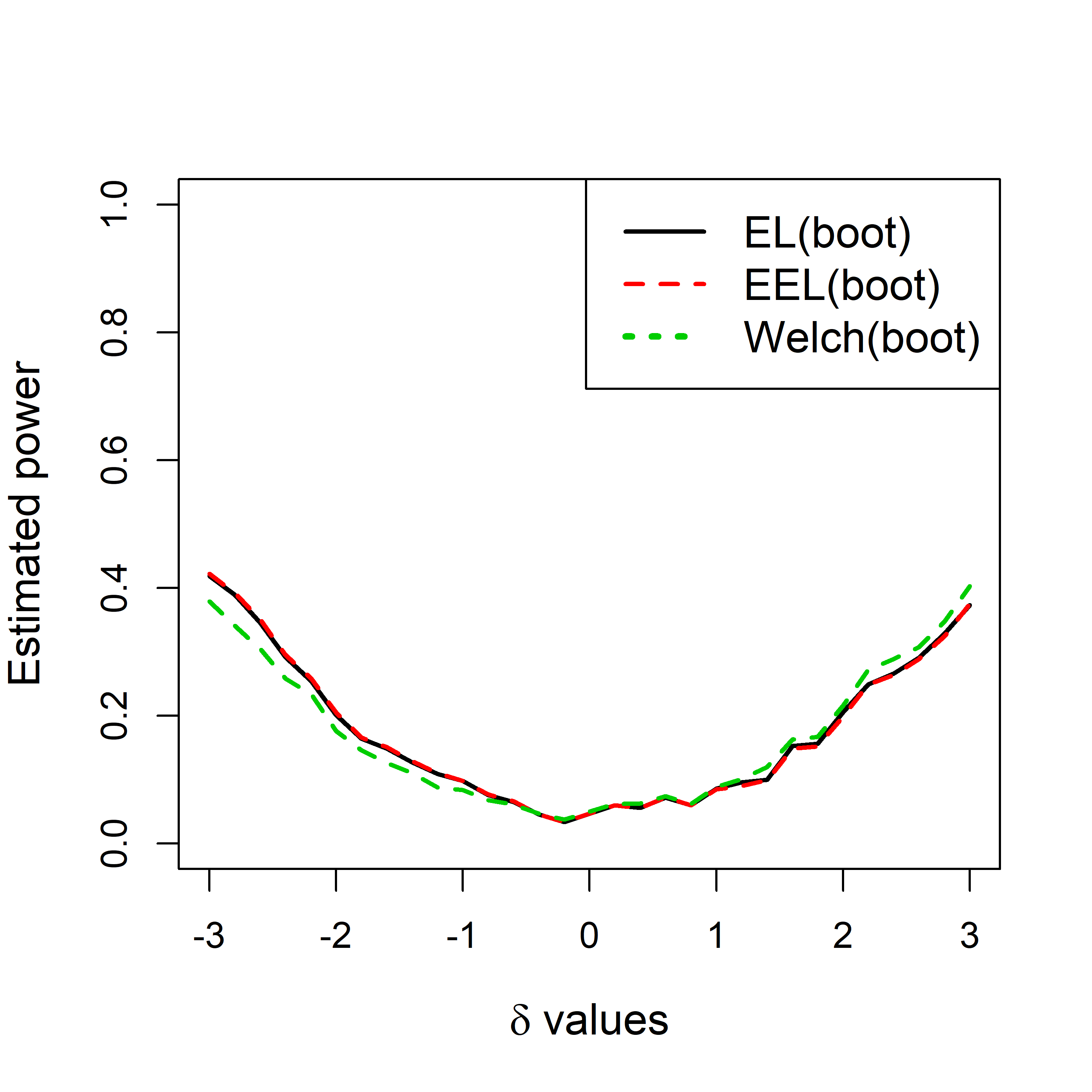}   \\
{\scriptsize (d)}   &   {\scriptsize (e)}  &   {\scriptsize (f)}   \\
\includegraphics[scale = 0.45, trim = 30 30 20 5]{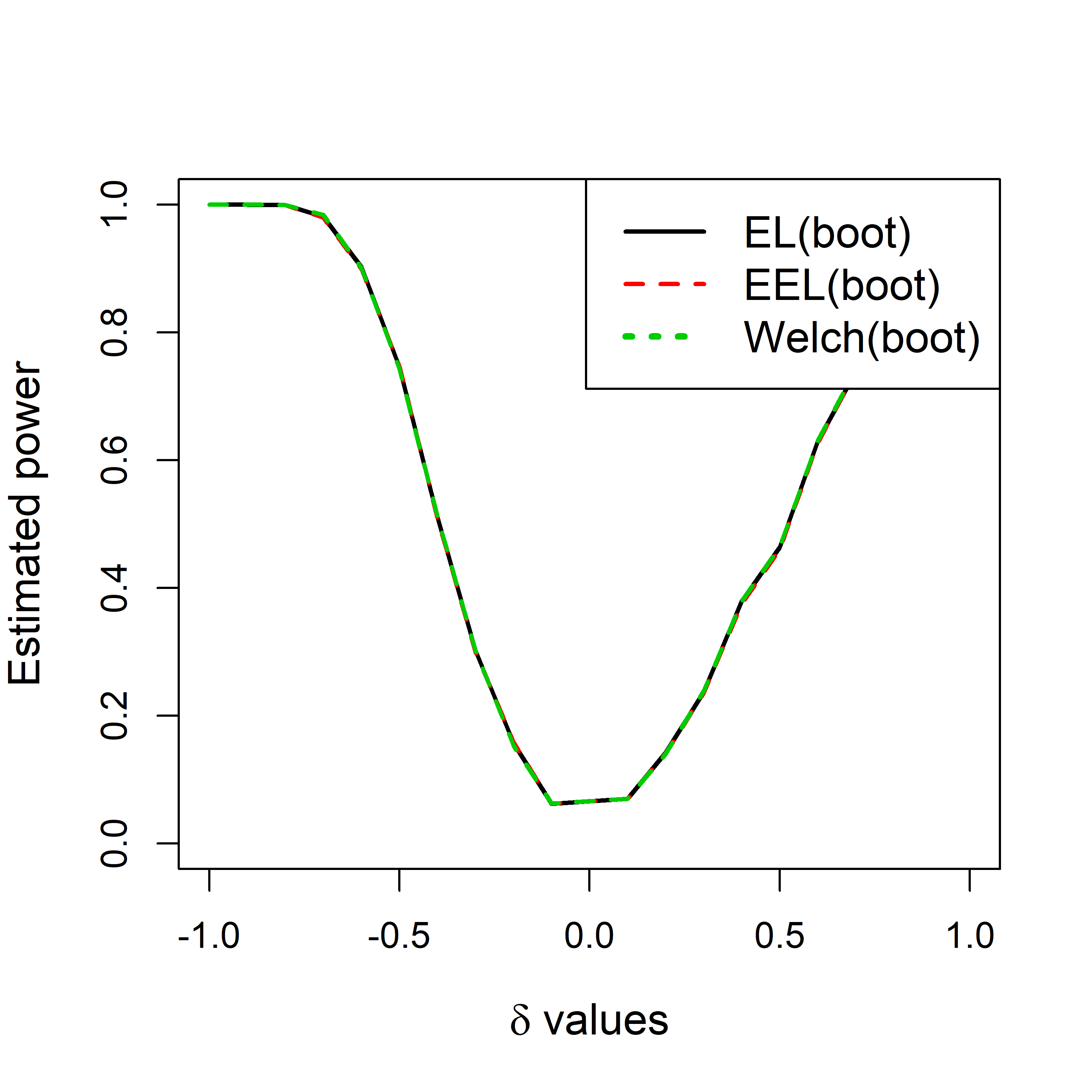} &
\includegraphics[scale = 0.45, trim = 30 30 0 5]{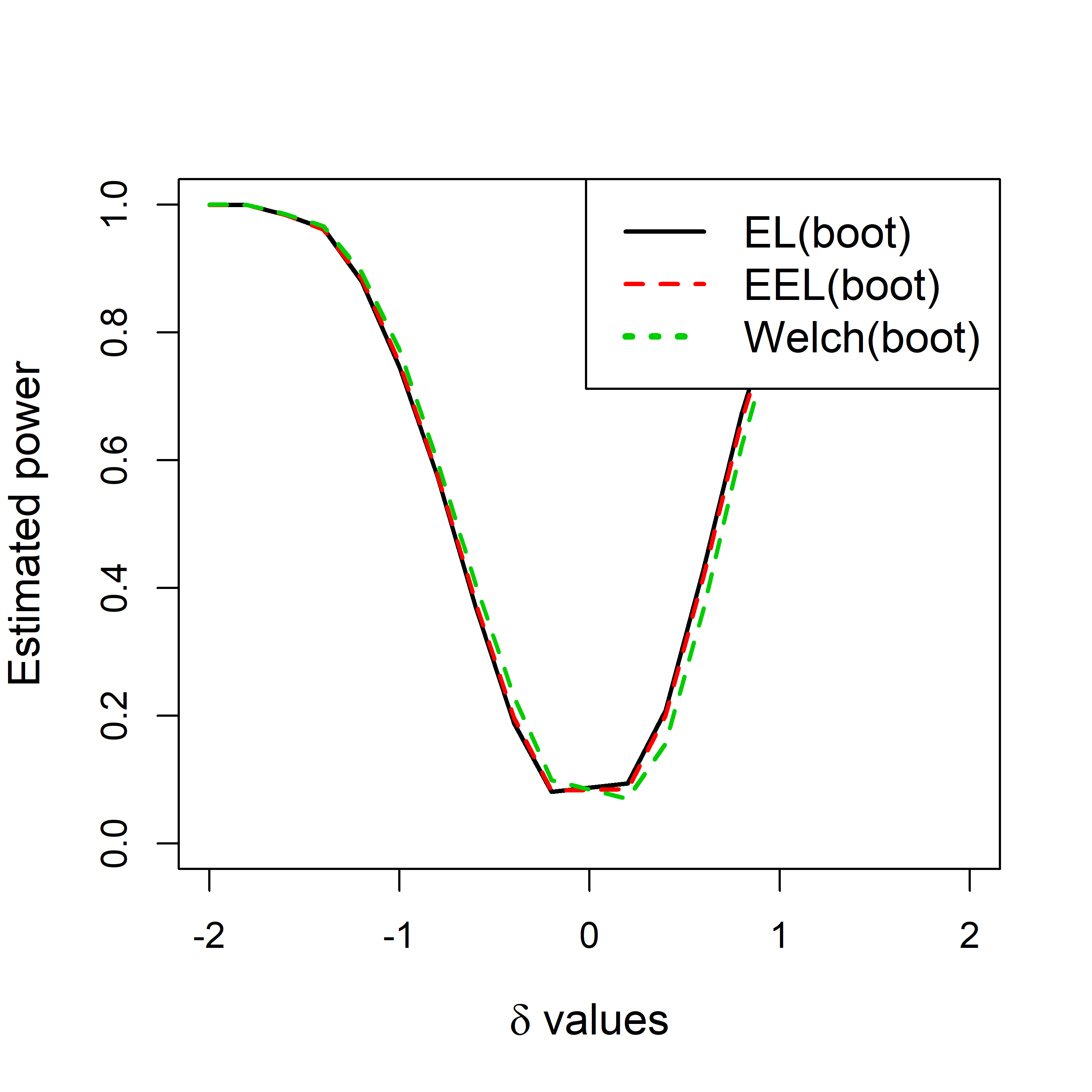} &
\includegraphics[scale = 0.45, trim = 40 30 20 5]{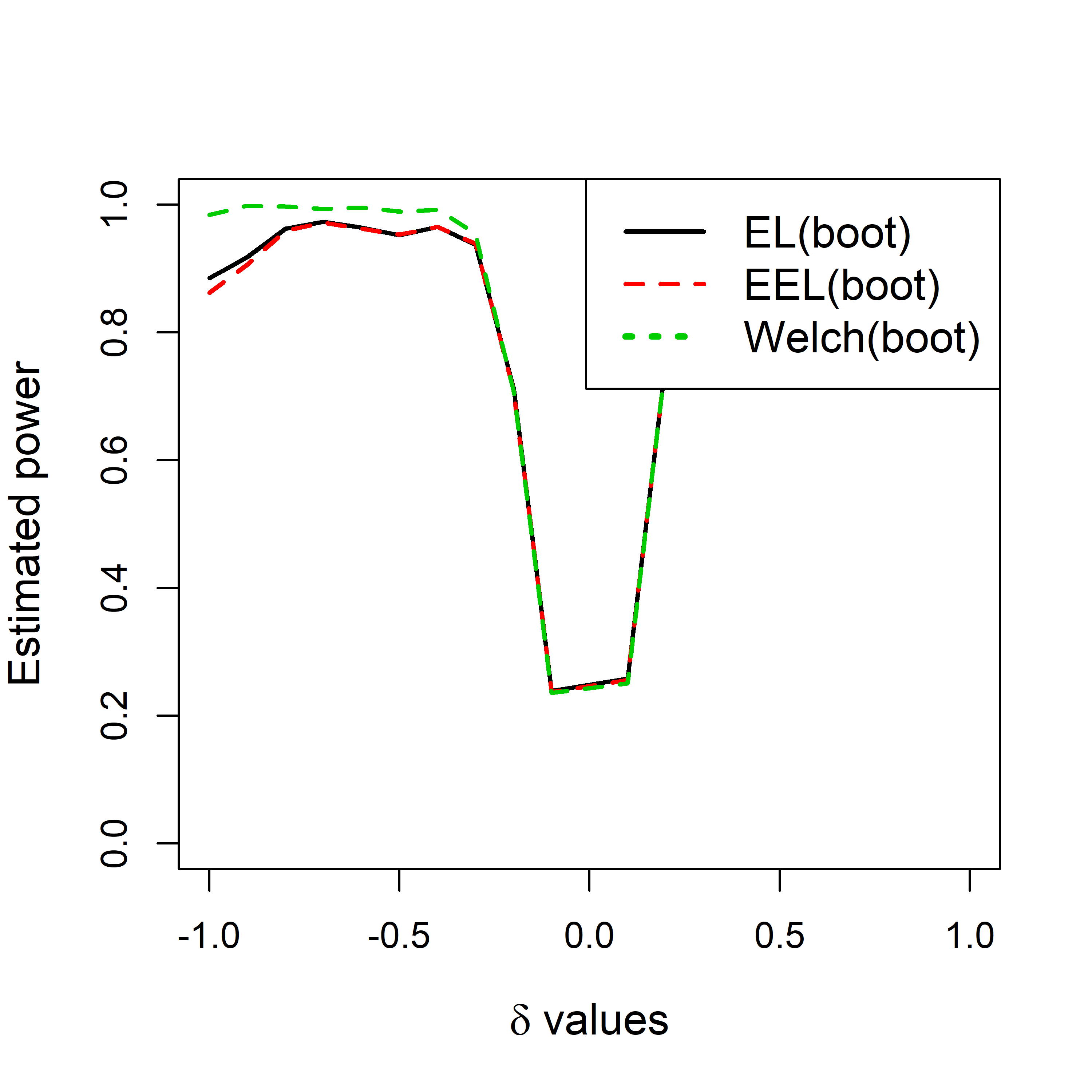}  \\
{\scriptsize (g)}   &   {\scriptsize (h)}  &   {\scriptsize (i)}   \\
\includegraphics[scale = 0.45, trim = 40 30 20 5]{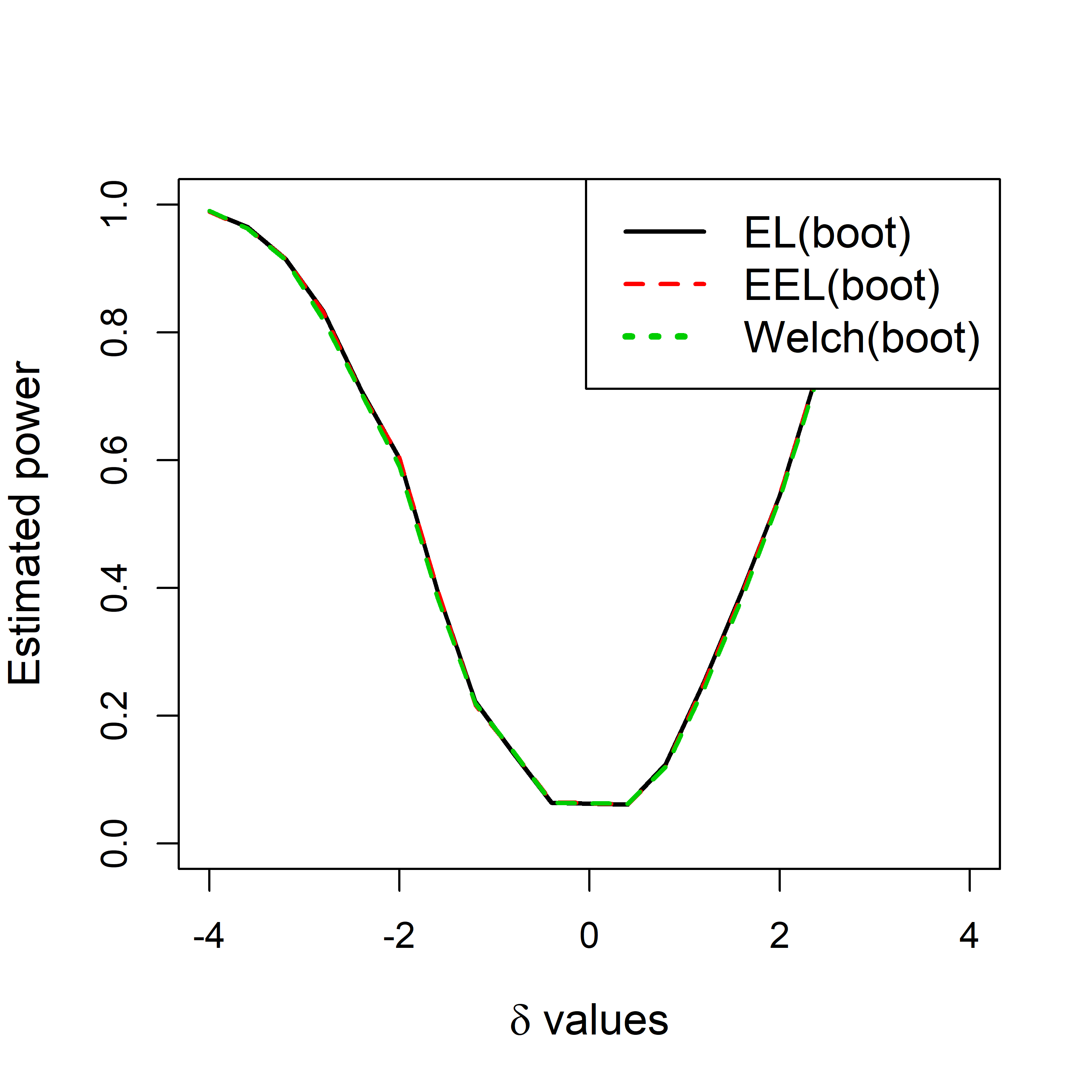} &
\includegraphics[scale = 0.45, trim = 30 30 20 5]{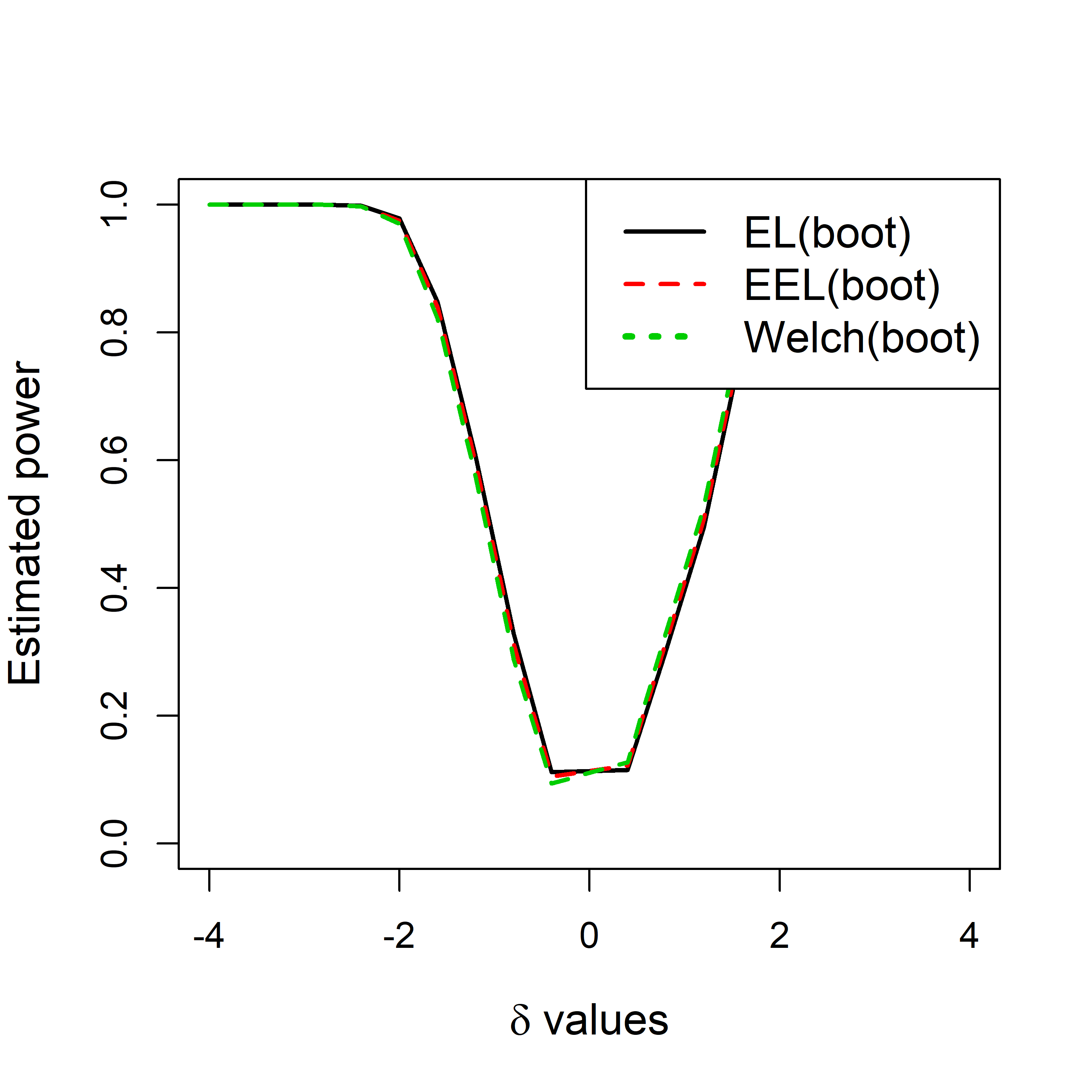} &
\includegraphics[scale = 0.45, trim = 30 30 0 5]{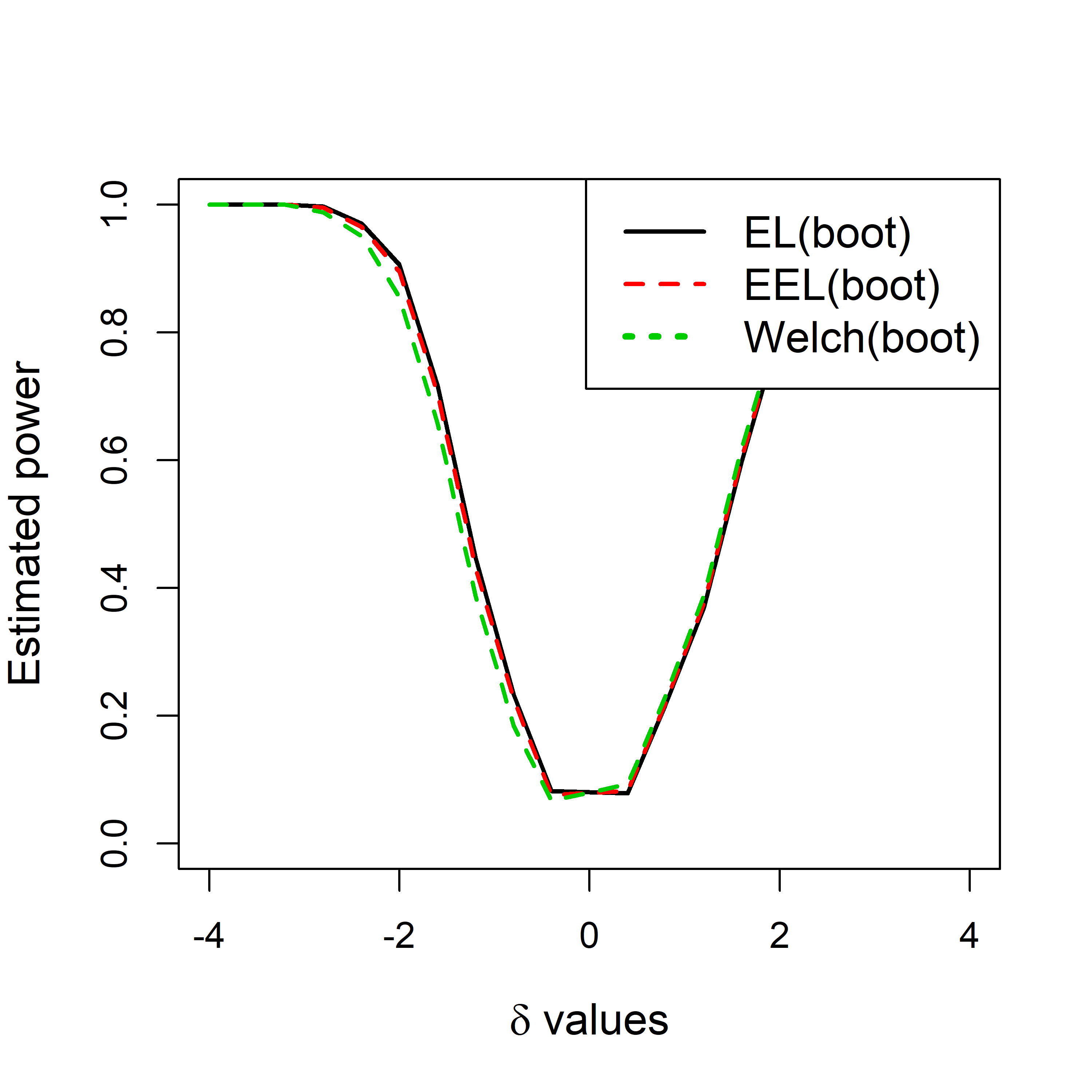}   \\
{\scriptsize (j)}   &   {\scriptsize (k)}  &   {\scriptsize (l)}   \\
\end{tabular}}
\caption{\textbf{Estimated power of EL, EEL and Welch, with bootstrap calibration applied to all three, for each of the 12 distributional scenarios}.}
\label{fig.powers} 
\end{figure}

Only the bootstrap calibrated procedures are considered, because the testing procedures had to be applied on size correct tests. Among them only the cases with the largest sample sizes $(50, 100)$ are presented due to space limitations. Figure \ref{fig.powers} manifests that the Welch $t$-test, the EL and the EEL testing procedures exhibit similar power levels.

\subsection{Asymptotic, bootstrap and exact p-values of the Welch $t$-test with very small sample sizes}
Detecting significant differences with small sample sizes, such as 10 observations per sample, can be really difficult and for this, only the type I error will be examined. The Welch $t$-test was applied in the 12 previously described scenarios and the asymptotic p-value, the bootstrap p-value and the exact\footnote{The term \textit{exact} stems from computing the p-value based on all possible permutations.} p-value are reported in Table \ref{exact}. When the sample sizes are very small, for example 10, the asymptotic p-value is shown to perform better that the exact and the bootstrap p-value. The estimated type I error based on the asymptotic p-value, was outside the permissible limits only once, twice based on the bootstrap p-value, while half of the times based on the exact p-value. 

\begin{table}[htp]
\caption{\textbf{Estimated type I error of the Welch $t$-test using the asymptotic p-value, bootstrap p-value and the exact p-value for all 12 scenarios when both samples had sizes equal to 10}. The numbers in bold indicate that the estimated probability was within the acceptable limits $\left(0.0365,0.0635\right)$. The first row refers to the method the p-value was computed.}
\begin{center}
{\begin{tabular}{l|ccc|l|ccc}  \\ \hline \hline
              & Asymptotic  &  Bootstrap  &  Exact  &  & Asymptotic  &  Bootstrap        &  Exact
              \\ \hline
Scenario (a)  &  \textbf{0.056}  &  0.068      &  0.077  &  Scenario (b)  &  \textbf{0.055}  &  \textbf{0.058}  &  0.067  \\
Scenario (c)  &  \textbf{0.053}  &  \textbf{0.053}   &  \textbf{0.057}  &  Scenario (d)  &  \textbf{0.043}  &  \textbf{0.044}  &  \textbf{0.059}  \\
Scenario (e)  &  \textbf{0.052}  &  \textbf{0.053}   &  \textbf{0.054}  &  Scenario (g)  &  0.023  &  0.023  &  0.026  \\
Scenario (g)  &  \textbf{0.054}  &  \textbf{0.058}   &  0.064  &  Scenario (h)  &  \textbf{0.054}  &  \textbf{0.062}  &  \textbf{0.058}  \\
Scenario (i)  &  \textbf{0.055}  &  \textbf{0.063}   &  \textbf{0.060}  &  Scenario (j)  &  \textbf{0.045}  &  \textbf{0.050}  &  \textbf{0.050}  \\
Scenario (k)  & \textbf{0.053}   &  \textbf{0.055}   &  0.073  &  Scenario (l)  &  \textbf{0.054}  &  \textbf{0.058} &  0.064  
\\ \hline \hline
\end{tabular}}
\end{center}
\label{exact}
\end{table}

\section{Examination of the computational cost using real gene expression data} \label{genes}

A third level of comparison is the computational time required for all the procedures. The latter is a highly essential feature in the era of massive and big data and in the field of Bioinformatics, and especially in differential gene expression analysis applied on microarrays or RNA-seq data, where the number of tests performed is in the order of tens of thousands.

One of the most common tasks in the field of Biology is to identify properties that are linked to specific phenotypes. A fundamental process that affects the biological behaviour of the cell is the expression levels of its genes. Discovering the genes that are expressed in a different abundance between two phenotypes has led to the identification of properties that are linked to specific conditions, like diseases. For this reason, experiments that measure the expression levels between two conditions, disease or no disease for example, are widely applied, leading to the construction of high-dimensional gene expression data, that include the information of the condition from which they derive. These experiments, mainly based on DNA microarray \citep{wheelan2008} or RNA-seq \citep{denoeud2008annotating} techniques, for measuring the expression levels, are typically returning datasets that consist of $55,000$\footnote{When the data come from humans or mice.} variables (genes, probesets) and a few tens of observations. A gene is defined as differentially expressed if a statistically significant mean difference is observed in the expression levels between the two experimental conditions.

Four gene expression datasets (GSE annotated)\footnote{From a biological point of interest, the data have been uniformly pre-processed, curated and automatically annotated.} downloaded from \href{http://dataome.mensxmachina.org/}{BioDataome} \citep{lakiotaki2018} were used to highlight the computational efficiency of the Welch $t$-test. The computational cost of all testing procedures, except for the WMW test, because of its poor performance in the simulation studies, was evaluated by computing the time (in seconds) required by each testing procedure to perform tens of thousands of hypothesis tests. The execution times appear on Table \ref{times}. 

Bootstrap calibration, with $499$ re-samples, was only applicable to the Welch $t$-test due to the excessively high computational cost required by the EEL and EL. $54,675$ Welch $t$-tests required centiseconds, and hundreds of seconds when bootstrap calibration was applied. It becomes obvious that EEL and EL with bootstrap calibration would require a few thousands of seconds and a few hours respectively. The Welch $t$-test with the new efficient bootstrap implementation \citep{chatzipantsiou2019} was only 3 times slower than EEL and 7 times faster than EL. 

\begin{table}[htp]
\caption{\textbf{Computational time (in seconds) required by each method}. 
The calibration technique, if one was applied is mentioned inside the parentheses. The $\chi^2$ refers to the $\chi^2$ distribution, "boot" refers to bootstrap calibration and Welch(boot$^*$) stands for the efficient bootstrap methodology for the Welch $t$-test \citep{chatzipantsiou2019}. The execution time of the fastest method appears in bold. The last row contains the total (for all 4 datasets) execution time required by each testing procedure. All functions were self-written in R with no C++ implementation of any test.}
\begin{center}
{\begin{tabular}{lll|cccccc} \\ \hline \hline
          &              &               & \multicolumn{5}{c}{\underline{Testing procedures}} \\
Datasets  & Sample sizes & \# Probesets  & EL($\chi^2$) & EEL($\chi^2$) &  Welch &  Welch(boot) & Welch(boot$^*$) \\ \hline
GSE12276  & (102, 102) &  54,675  &  126.60  &  6.41  &  \textbf{0.12}  &  1046.78  &  17.31  \\
GSE50081  & (91, 90)   &  54,675  &  128.75  &  5.50  &  \textbf{0.12}  &  1167.10  &  15.67  \\
GSE62452  & (65, 65)   &  33,297  &  55.87   &  2.84  &  \textbf{0.06}  &  727.09   &  7.28   \\
GSE28735  & (45, 45)   &  33,297  &  54.44   &  2.41  &  \textbf{0.07}  &  708.09   &  7.03   \\  \hline 
Total     &            &          &  365.66  &  17.16 &  \textbf{0.37}  &  3649.06  &  47.29  \\  \hline \hline
\end{tabular}}
\end{center}
\label{times}
\end{table}

\section{Conclusions} \label{conclusions}
In most cases examined, the nominal level of the type I error was not attained by the non-parametric testing procedures unless the sample sizes were relatively large. Further, the calibration of the EL and EEL test statistics using the $t$ distribution proved successful. Improved accuracy in estimating the type I error compared to the $\chi^2$, especially for EL was found. Bootstrap calibration, on the other hand, performed satisfactorily in nearly all cases, especially for the small sample size cases, with the computational burden being the price to pay. The Welch $t$-test on the contrary is numerical optimization free and applying the efficient bootstrap methodology of \cite{chatzipantsiou2019} increases the computational cost only slightly. With relatively larger sample sizes though, Welch $t$-test requires no bootstrap calibration as it is size correct and furthermore it is computationally extremely efficient. Surprisingly enough, with very small sample sizes ($n=10$), the Welch $t$-test attained the type I error showing that bootstrap calibration is not always necessary. We found that computation of the exact p-value can lead to wrong conclusions. The WMW test was manifested to be highly inaccurate in terms of type I error, even if the exact p-value was calculated. In addition the exact p-value cannot be computed when ties are present in the data, a usual phenomenon with discrete data. Both arguments showed that WMW should not be considered as a competing non-parametric alternative to the Welch $t$-test.

A third disadvantage of EL and EEL, not previously discussed, is related to the convex hull of the combined samples. When the two samples are well separated there is no intersection between them, and hence EL and EEL have no solution and EL could yield negative $p_i$s. The Welch $t-$test on the contrary is free from this limitation, that commonly appears in higher dimensions. 

An interesting discovery is the robustness of the Welch $t$-test to different types of data (continuous, discrete, constrained or unconstrained support) and to distributional deviations from normality. This is supported by the central limit theorem that is valid even for relatively small sample sizes. Our findings suggest that the use of Welch $t$-test, with or without bootstrap calibration, should be preferred over EL, EEL, and the WMW test, for comparing two population means. Additional advantages of the Welch $t$-test include its computational efficiency, its simplicity in programming and interpretation as well as the fact that it is free of any convex hull limitations. 




\appendix

\section{Proof of the EL test statistic} \label{A1}
Let us denote for brevity and convenience purposes the 2 samples\footnote{The generalisation to more samples is straightforward.} by ${\bf X}_1$ and ${\bf X}_2$. The 2 constraints are
\begin{eqnarray*} 
f_j(\lambda_j)=\frac{1}{n_j}\sum_{i=1}^{n_j}\left\lbrace\left[1+\lambda_j\left(x_{ji}-\mu \right)\right]^{-1}\left(x_{ji}-\mu\right)\right\rbrace=0 \ \ (j=1,2).
\end{eqnarray*}
The first derivative of $f_j(\lambda_j)$ with respect to $\lambda_j$ is
\begin{eqnarray*}
\frac{\partial f_j\left(\lambda_j\right)}{\partial\lambda_j} = -\frac{1}{n_j}\sum_{i=1}^{n_j}\left\lbrace\left[1+\lambda_j\left(x_{ji}-\mu\right)\right]^{-2}\left(x_{ji}-\mu\right) \left(x_{ji}-\mu\right) \right\rbrace.
\end{eqnarray*}
For each of these 2 constraints, the Maclaurin expansion for the vector $\lambda_j$ is applied and by keeping the leading term only, $f_j\left(\lambda_j \right)$ can be written as 
$\lambda_j = s^2_j\left(\mu\right)^{-1}\left(\bar{x}_j-\mu\right)+O_{p}\left(n_0^{-1/2}\right),$
where $n_0=\min\left\lbrace n_1, n_2\right\rbrace$. The log-likelihood ratio test statistic becomes
\begin{eqnarray*}
\Lambda =2\sum_{j=1}^2\sum_{i=1}^{n_j}\log{\left[1+\lambda_j\left(x_{ji}-\mu\right)\right]}=
\sum_{j=1}^2g_j\left(\lambda_j\right). 
\end{eqnarray*}
The first and second derivatives of $g_j(\lambda_j)$ with respect to $\lambda_j$ are
\begin{eqnarray*}
\frac{\partial g_j\left(\lambda_j \right)}{\partial\lambda_j} &=& 2\sum_{i=1}^{n_j}\left\lbrace\left[1+\lambda_j\left(x_{ji}-\mu\right)\right]^{-1}\left(x_{ji}-\mu\right) \right\rbrace \\
\frac{\partial^2 g_j\left(\lambda_j \right)}{\partial\lambda_j^2} &=& 2\sum_{i=1}^{n_j}\left\lbrace\left[1+\lambda_j\left(x_{ji}-\mu\right)\right]^{-2}\left(x_{ji}-\mu\right)  \left(x_{ji}-\mu\right) \right\rbrace \\
\end{eqnarray*}
and thus $\Lambda$ becomes 
$
\Lambda = \sum_{j=1}^2 \frac{n_j\left(\bar{x}_j-\mu\right)^2}{s^2_j\left(\mu\right)}+O_{p}\left(n_0^{-1}\right).
$

\section{Proof of the EEL test statistic} \label{A2}
Here, the two sample means case but in a simpler form will be presented, using a similar approach to \cite{jing1997} . The three constraints are
\begin{eqnarray*}
\begin{array}{ccc}
\left(\sum_{k=1}^{n_j}e^{\lambda_j x_{kj}}\right)^{-1}\left(\sum_{i=1}^{n_j} x_{ji}e^{\lambda_j
x_{ji}}\right) -\mu &=& 0  \ \ (j=1,2) \\
n_1\lambda_1+n_2\lambda_2 &=& 0.
\end{array}
\end{eqnarray*}
Equating the first 2 constraints and by using the third constraint the following holds:
\begin{eqnarray*} 
\left(\sum_{j=1}^{n_1}e^{\lambda x_{1j}}\right)^{-1}\left(\sum_{i=1}^{n_1} x_{1i}e^{\lambda x_{1i}}\right) =
\left(\sum_{j=1}^{n_2}e^{-\frac{n_1}{n_2}\lambda x_{2j}}\right)^{-1}\left(\sum_{i=1}^{n_2} x_{2i}e^{-\frac{n_1}{n_2}\lambda x_{2i}}\right). \\
\end{eqnarray*}
This trick allows to avoid the estimation of the common mean. Assuming $2$ samples and $\mu$ being the common mean as before, the probabilities become
\begin{eqnarray*}  
p_{ji}=\left(\sum_{m=1}^{n_j}e^{\lambda_j x_{jm}}\right)^{-1}e^{\lambda_j x_{ji}} = \left[\sum_{m=1}^{n_j}e^{\lambda_j\left(x_{jm}-\mu\right)}\right]^{-1}e^{\lambda_j\left( x_{ji}-\mu\right)}\ \ \  \left( j=1,2 \right).
\end{eqnarray*}
The constraints for the EEL are
\begin{eqnarray*} 
f(\lambda_j) = \left(\sum_{m=1}^{n_j}e^{\lambda_j x_{jm}}\right)^{-1}\left(\sum_{i=1}^{n_j} x_{ji}e^{\lambda_j x_{ji}}\right)-\mu = 0 \ \ \left( j=1,2 \right). 
\end{eqnarray*}
The first derivative  of $f_j\left(\lambda_j\right)$ with respect to $\lambda_j$ is
\begin{eqnarray*}
\frac{\partial{f}_j\left(\lambda_j\right)}{\partial\lambda_j} = -\left(\sum_{m=1}^{n_j}e^{\lambda_j x_{jm}}\right)^{-2} \left(\sum_{i=1}^{n_j} x_{ji}e^{\lambda_j x_{ji}}\right)^2
+\left(\sum_{m=1}^{n_j}e^{\lambda_j x_{jm}}\right)^{-1} \sum_{i=1}^{n_j} x^2_{ji}e^{\lambda_j x_{ji}}  
\end{eqnarray*}
Note that here the constraint can be written as $\sum_{i=1}^np_i x_i-\mu=0$ as this makes no difference, whereas in the EL the constraint must stay $\sum_{i=1}^np_i\left(x_i-\mu\right)=0$, since the resulting probabilities in the EL depend upon $\mu$. Under $H_0$ the value of each of the $\lambda_j$ using McLaurin series becomes
$\lambda_j = -s^2_j\left(\bar{x}_j\right)^{-1}\left(\bar{x}_j-\mu\right)+o_{p}\left(n_j^{-1/2}\right),$
and the log-likelihood ratio test statistic is 
\begin{eqnarray*} 
& & \Lambda = -2\sum_{j=1}^2\sum_{i=1}^{n_j}\log{\left[n_j\left(\sum_{m=1}^{n_j}e^{\lambda_j x_{jm}}\right)^{-1}e^{\lambda_j x_{ji}}\right]} = 2\sum_{j=1}^2g_j\left(\lambda_j\right), \\
& & \text{were} \ \  g_j\left(\lambda_j\right) = n_j\log{\sum_{i=1}^{n_j}e^{\lambda_j x_{ji}}}-\sum_{i=1}^{n_j}\lambda_j x_{ji}-\sum_{i=1}^{n_j}\log{n_j} \nonumber
\end{eqnarray*}
By using the first and second derivative of $g_j\left(\lambda_j \right)$ with respect to $\lambda_j$ 
and keeping the two leading terms in the MacLaurin series, ($\Lambda$) can be written as 
\begin{eqnarray*}
\Lambda=\sum_{j=1}^2n_j\frac{\left(\bar{x}_j-\mu\right)^2}{s^2_j\left(\bar{x}_j\right)}+O_{p}\left(n_0^{-1/2}\right).
\end{eqnarray*}


\end{document}